\documentclass[12pt,dvips]{article}

\usepackage{rotating}
\usepackage{hhline}
\usepackage{array}
\usepackage{epsfig}
\usepackage{amssymb}

\begin{document}
\tolerance=100000

\newcommand{\imag}{\Im {\rm m}}
\newcommand{\real}{\Re {\rm e}}

\def\tablename{\bf Table}%
\def\figurename{\bf Figure}%

\newcommand{\sts}{\scriptstyle}
\newcommand{\ngs}{\!\!\!\!\!\!}
\newcommand{\rb}[2]{\raisebox{#1}[-#1]{#2}}
\newcommand{\CP}{${\cal CP}$~}
\newcommand{\sbomu}{\frac{\sin 2 \beta}{2 \mu}}
\newcommand{\kmol}{\frac{\kappa \mu}{\lambda}}
\newcommand{\s}{\\ \vspace*{-3.5mm}}
\newcommand{\lsim}{\raisebox{-0.13cm}{~\shortstack{$<$\\[-0.07cm] $\sim$}}~}
\newcommand{\gsim}{\raisebox{-0.13cm}{~\shortstack{$>$\\[-0.07cm] $\sim$}}~}

\begin{titlepage}

\begin{flushright}
DESY 04-088\\
Edinburgh 2004/07\\
KIAS-P04026\\[1mm]
\today
\end{flushright}

\vspace{1cm}

\begin{center}
{\Large \bf The Neutralino Sector of the           \\[2mm]
            Next-to-Minimal Supersymmetric
            Standard Model}\\[1cm]
{\large S.Y. Choi$^{1}$, D.J.~Miller$^{2}$ and P.M.~Zerwas$^{3}$}\\[1cm]
{\it $^1$ Physics Department, Chonbuk National University, Chonju 561-756,
          Korea\\
     $^2$ School of Physics, The University of Edinburgh, Edinburgh EH9 3JZ,
          Scotland\\
     $^3$ Deutsches Elektronen--Synchrotron DESY, D--22603 Hamburg, Germany}\\
\end{center}

\renewcommand{\thefootnote}{\fnsymbol{footnote}}
\vspace{3cm}

\begin{abstract}
\noindent The Next--to--Minimal Supersymmetric Standard Model (NMSSM)
includes a Higgs iso-singlet superfield in addition to the two Higgs
doublet superfields of the minimal extension. If the Higgs
fields remain weakly coupled up to the GUT scale, as naturally
motivated by the concept of supersymmetry, the mixing between singlet
and doublet fields is small and can be treated perturbatively. The
mass spectrum and mixing matrix of the neutralino sector can be
analyzed analytically and the structure of this 5--state system is
under good theoretical control. We also determine decay modes and
production channels in sfermion cascade decays to these particles at
the LHC and pair production in $e^+e^-$ colliders.
\end{abstract}

\end{titlepage}

\renewcommand{\thefootnote}{\fnsymbol{footnote}}

\section{Introduction}

The Minimal Supersymmetric Standard Model (MSSM) \cite{MSSMa,MSSMb}
opens the path to the analysis of supersymmetric theories.  Arguments
have been advanced however that suggest extensions beyond this minimal
version. One well--motivated example is the Next--to--Minimal
Supersymmetric Standard Model (NMSSM) \cite{NMSSM} in which an iso--singlet
Higgs superfield $\hat{S}$ is introduced in addition to the two iso--doublet
Higgs fields $\hat{H}_{u,d}$ incorporated in the MSSM to generate
electroweak symmetry breaking. Such an extension offers a possible solution
of the $\mu$ problem, generating in a natural way, a value of the order of
the electroweak breaking scale $v$; this is achieved by identifying $\mu$,
apart from the ${\cal O}(1)$ coupling, with the vacuum expectation value of
the scalar component $S$ of the new iso--singlet field. [For a recent summary
of this construct see Ref.\cite{mnz}; a useful code has been made available
in Ref.\cite{ref:code}.] \s

The superpotential of the NMSSM includes, besides the usual MSSM $W_Y$
Yukawa components, an additional term, which couples the iso--singlet
to the two iso--doublet Higgs fields, plus the self--coupling of the
iso--singlet:
\begin{eqnarray}
W=W_Y +\lambda \hat{S}(\hat{H}_u \hat{H}_d)+\frac{1}{3}\kappa\hat{S}^3
\label{eq:superpotential}
\end{eqnarray}
The two parameters $\lambda$ and $\kappa$ are dimensionless. By demanding
the Higgs fields remain weakly interacting up to the GUT scale, the two
couplings are bounded at the electroweak scale by the inequalities
$\lambda, \kappa \lsim 0.7$. While the scalar Higgs sector includes several
soft supersymmetry breaking parameters, the Lagrangian of the gaugino/higgsino
sector is complemented only by the familiar SU(2) and U(1) gaugino mass terms.
As a result, the parameter space of the neutralino sector is much less
complex than the Higgs space.\s

The superpotential without the singlet self--coupling, {\it i.e.}
$\kappa=0$, incorporates a Peccei--Quinn (PQ) symmetry:
$\{\hat{H}_u(1),\, \hat{H}_d(1),\, \hat{S}(-2),\, \hat{Q}(-1),\,
\hat{U}(0),\, \hat{D}(0),\, \hat{L}(-1),\, \hat{E}(0)\}$.  $\hat{Q}$
and $\hat{L}$ are the quark and lepton SU(2) doublet superfields,
while $\hat{U}$, $\hat{D}$ and $\hat{E}$ are the up-- and down--quark
and lepton SU(2) singlet superfields respectively. The integer of each
parenthesis indicates the PQ charge of the corresponding
superfield. The spontaneous breaking of this symmetry by the non--zero
vacuum expectation value $v_s/\sqrt{2}$ of the scalar $S$ field gives
rise to a massless Goldstone boson.
However, when $\kappa \neq 0$, the mass is lifted to a non--zero value by
the self--interaction of the $S$ field. Still, a discrete $\mathbb{Z}_3$
symmetry is left which would lead to the formation of domain walls in
the early Universe. This problem can be tamed by
introducing new interactions of the inverse Planck size that, however, do
not affect the low--energy effective NMSSM
theory \cite{tadpoles}.\s

In contrast to the Higgs sector, masses and mixings in the chargino system
are not affected by the singlet extension. [Of course new decays such as
$S\rightarrow \tilde{\chi}^+\tilde{\chi}^-$ or $\tilde{\chi}^+\rightarrow
\tilde{\chi}_5 H^+$ may be possible if allowed kinematically.]\s

So far the {\it supersymmetric} particle spectrum of the NMSSM has
received only little attention in the NMSSM literature,
Refs.\cite{NMSSM}, \cite{nmssm1}-\cite{ref:EH}. In this report we
attempt a {\it systematic analytical analysis of the neutralino
system}.  In contrast to the MSSM where exact solutions of the mass
spectrum and mixing parameters can be constructed mathematically in
closed form, this is not possible any more for the NMSSM in which the
eigenvalue equation is of 5th order, not allowing closed
solutions. However, since the coupling between singlet and doublet
fields is weak, $\lambda v/\sqrt{2}\sim O(10^2)\, {\rm GeV}$, compared
with the typical supersymmetry scale $M_{1,2}$ and $\mu=\lambda
v_s/\sqrt{2}\sim O(10^3)\, {\rm GeV}$, a perturbative expansion of the
solution gives rise to a good approximation of the mass spectrum while
the magnitude of the matrix elements in the mixing matrix is at least
qualitatively well understood. The usefulness of a perturbative
expansion has also been noticed in Ref.\cite{ref:EH}; however, here,
extending the Higgs analysis
in Ref.\cite{mnz}, we work out this approach systematically for all
facets of the NMSSM.\s

While $\kappa$ plays a crucial r\^{o}le in the Higgs sector, it is
less crucial for the neutralino system. The size of $\kappa v_s$, with
$v_s\lsim 15 \, v$ to maintain a link with the electroweak scale, just
determines the singlino mass before modified by mixing effects.  Once
masses and mixings are determined, the couplings of the neutralinos to
the electroweak gauge bosons and to scalar/fermionic matter particles
are fixed. Decay widths and production rates of the five neutralinos
can subsequently be predicted for squark cascades at the LHC \cite{LHC}
and $e^+e^-$ annihilation at prospective linear colliders \cite{FLC}.\s

The report is organized as follows. In Sect.~\ref{sec:sec2} we
describe the neutralino sector of supersymmetric models in which
the pair of Higgs doublet superfields is augmented by an additional
iso-singlet field.  In Sect.~\ref{sec:sec3} we show how, for a naturally
expected weak coupling, the properties of the four standard neutralinos are
modified; moreover the properties of the fifth neutralino, the new
singlino--dominated state, are calculated. All these spectra and
mixings are pre--determined analytically before the surprisingly
good quality of the weak--coupling expansion is demonstrated by
comparison with numerical solutions. In this way we achieve a
satisfactory theoretical understanding of the system. In the limit
of large gaugino mass parameters $M_{1,2}$ compared with the
higgsino mass parameter $\mu$, or {\it vice versa}, the MSSM part
can be easily diagonalized analytically and a clear and simple
picture of the entire system emerges. The section is concluded by
a lovely toy model in which we set $M_1=M_2$ and $\tan\beta=1$;
this set allows us to solve the system exactly, leading to
transparent closed expressions for the neutralino mass spectrum
and the mixing parameters. A sample of decay widths and production
cross sections for the neutralinos is presented in Sect.~\ref{sec:sec4}.
The results are summarized in Sect.~\ref{sec:conclusion} and technical
details of the diagonalization procedure for the $5\times 5$ neutralino
mass matrix are described in the Appendix.

%=============================================
\section{The NMSSM Neutralino Sector}
\label{sec:sec2}
%=============================================

\subsection{The NMSSM neutralino mass and mixing matrix}

The Lagrangian of the neutralino system can be derived from the superpotential
defined in Eq.(\ref{eq:superpotential}), complemented by the SU(2) and
U(1) mass terms in the soft supersymmetry breaking Lagrangian. After
breaking the [electroweak] symmetry spontaneously by introducing
non--zero vacuum expectation values of the iso-doublet and singlet
Higgs fields,
\begin{eqnarray}
\langle H_{u} \rangle = \frac{1}{\sqrt{2}} \sin \beta\,
\left( \begin{array}{c} 0 \\ v \end{array} \right),\qquad
\langle H_{d} \rangle = \frac{1}{\sqrt{2}} \cos \beta\,
\left( \begin{array}{c} v \\ 0 \end{array} \right), \qquad
\langle S \rangle = v_s/\sqrt{2}
\label{eq:vevs}
\end{eqnarray}
the Higgs--higgsino mass parameter
\begin{eqnarray}
\mu = \lambda v_s/\sqrt{2}
\end{eqnarray}
is generated and, subsequently, the neutralino mass matrix
\begin{eqnarray*}
{\cal M}_5 = \left(\begin{array}{cc}
                    {\cal M}  & X \\
                      X^T  & \mu_\kappa
                    \end{array}\right)
\end{eqnarray*}
with a hierarchical structure as analyzed in the Appendix,
can be written in detail as:
\begin{eqnarray}
{\cal M}_5 =\left(\begin{array}{cccc|c}
  M_1  &   0   & -m_Z\, c_\beta\, s_W  &  m_Z\, s_\beta\, s_W  &  0  \\
   0   &  M_2  &  m_Z\, c_\beta\, c_W  & -m_Z\, s_\beta\, c_W  &  0  \\
-m_Z\, c_\beta\, s_W &  m_Z\, c_\beta\, c_W  &   0
                     & -\mu & -\mu_\lambda\,s_\beta\\
 m_Z\, s_\beta\, s_W & -m_Z\, s_\beta\, c_W  & -\mu
                     &   0  & -\mu_\lambda\,c_\beta\\[2mm]
 \cline{1-4}
  0    &   0   & -\mu_\lambda\,s_\beta & \multicolumn{1}{c}{-\mu_\lambda\,
   c_\beta}    & \mu_\kappa
                   \end{array}\right)
\label{eq:mass_matrix}
\end{eqnarray}
This $5\times 5$ mass matrix is constructed from the standard $4\times 4$ MSSM
neutralino mass matrix ${\cal M}$ in the upper left corner, the mass term of
the higgsino component $\tilde{S}$ of the singlet superfield $\hat{S}$,
\begin{eqnarray}
\mu_\kappa  = 2\kappa\, v_s/\sqrt{2}
\label{eq:mu_kappa}
\end{eqnarray}
and the mixing between doublets and singlet parameterized by
\begin{eqnarray}
\mu_\lambda = \lambda\,v/\sqrt{2}
\label{eq:mu_lambda}
\end{eqnarray}
The mass matrix ${\cal M}_5$ is defined in the group basis $(\tilde{B},
\tilde{W}^3, \tilde{H}^0_d, \tilde{H}^0_u, \tilde{S})$.  As usual,
$M_1$ and $M_2$ are the soft SUSY breaking U(1) and SU(2) gaugino mass
parameters, $\tan\beta$ is the ratio of the vacuum expectation values
of the two neutral SU(2) Higgs doublet fields (as defined in
Eq.~(\ref{eq:vevs})), $s_\beta=\sin\beta$, $c_\beta=\cos\beta$, and $s_W,
c_W$, $t_W$ are the sine, cosine and tangent of the electroweak mixing
angle $\theta_W$.\s

Since the neutralino mass matrix (\ref{eq:mass_matrix}) is symmetric
and real, it can be diagonalized by an orthogonal matrix $V^5$. The
mass eigenvalues are real but not necessarily positive. They can be
mapped onto positive values by supplementing the rotation matrix to
$N^5= \Phi^5 V^5$ with the diagonal phase matrix $(\Phi^5)_{kl}=1
(i)\, \delta_{kl}$ in case of positive (negative) eigenvalues so that
$N^{5*}\, {\cal M}_5\, N^{5\dagger}$ is positive diagonal. The
physical neutralino states $\tilde{\chi}^0_i$ $[ i = 1-4]$ are ordered
according to ascending mass values while $\tilde{\chi}^0_5$ is the
predominantly singlino state.\footnote{Note that the ordering of the
masses according to ascending values is accomplished easily after the
diagonalization process is finalized. For the intermediate steps it is
however convenient to use the indices $i=1,2,3,4$ for the former MSSM
type states and $i=5$ for the additional state originating from the
singlino field as suggested by the structure of ${\cal M}_5$ in Eq.(4).}
They are mixtures
\begin{eqnarray}
\tilde{\chi}^0_i
    = N^5_{ij}\, ( \tilde{B}, \tilde{W}^3, \tilde{H}_d,
                      \tilde{H}_u, \tilde{S})_j \qquad [i = 1-5]
\end{eqnarray}
of the U(1), SU(2) gauginos, the doublet higgsinos and the singlino.\s

The unitary matrix $N^5$ defines the couplings of the mass eigenstates
$\tilde{\chi}^0_i$  to other particles. For the neutralino
production processes it is sufficient to consider the
neutralino--neutralino--$Z$ vertices
\begin{eqnarray}
&& \langle \tilde{\chi}^0_{iL}|Z|\tilde{\chi}^0_{jL}\rangle
 = -\frac{g}{2 c_W}\left( N^5_{i3} N^{5*}_{j3}-N^5_{i4} N^{5*}_{j4}\right)
   \nonumber\\
&& \langle \tilde{\chi}^0_{iR}|Z|\tilde{\chi}^0_{jR}\rangle
 = +\frac{g}{2 c_W}\left( N^{5*}_{i3} N^5_{j3}-N^{5*}_{i4} N^5_{j4}\right)
\label{eq:coupling_set1}
\end{eqnarray}
and the fermion-sfermion-neutralino vertices
\begin{eqnarray}
&& \langle \tilde{\chi}^0_{iR}|\tilde{f}_L|f_L\rangle
 = -\sqrt{2}\frac{g}{c_W}
    \left[I^f_3 N^{5*}_{i2}c_W+(Q_f-I^f_3) N^{5*}_{i1} s_W\right]
    \nonumber\\
&& \langle \tilde{\chi}^0_{iL}|\tilde{f}_R|f_R\rangle
 =  2 g \, Q_f\, s_W N^5_{i1}
\label{eq:coupling_set2}
\end{eqnarray}
The coupling $g$ is the SU(2) gauge coupling, $I^f_3$ is the SU(2)
isospin 3-component and $Q_f$ is the electric charge of the fermion
$f$.  In Eq.~(\ref{eq:coupling_set2}) the coupling to the higgsino
component, which is proportional to the fermion mass, has been
neglected for ``light flavors''. The more involved Higgs couplings to the
neutralinos are listed in detail in Sect.~\ref{sec:sec4}. \s

\subsection{NMSSM parameter range}

In contrast to the Higgs sector only two additional parameters
$\lambda$ and $\kappa$ are introduced in the NMSSM neutralino sector
as compared to that of the MSSM including $\mu$. Assuming that the fields remain
weakly interacting up to the GUT scale, the two couplings are bounded
at the electroweak scale by the inequality
\begin{eqnarray}
\sqrt{ \lambda^2+\kappa^2}\, \lsim\, 0.7
\end{eqnarray}
Moreover, the renormalization group (RG) evolution of the
couplings points to $\kappa \lsim \lambda$ as preferential target
domain if the evolution starts from a random distribution of the
couplings $\kappa_U, \lambda_U\leq 2\pi$ at the GUT scale
\cite{mnz}.\s

While $v=\sqrt{v^2_u+v^2_d}=246$ GeV is fixed by the Fermi coupling $G_F$,
the parameter $v_s$ should be expected in the same range,
\begin{eqnarray}
v_s \lsim 15\, v
\end{eqnarray}
in compliance with the arguments for introducing the NMSSM. A RG
analysis of the entire set of parameters shows that a low value of
$\tan\beta$ is favored \cite{mnz}. Current experimental analyses of
$\tan\beta$ assume MSSM relations for the couplings; they are
modified in the NMSSM and the results in this extended scenario
are less restrictive. \s

Since the size of the doublet--singlet mixing is set by $\mu_\lambda =
\lambda v/\sqrt{2}$, the mixing interaction is expected to be
small\footnote{$\mu_{\lambda}$ is expected to have a lower limit from
cosmological arguments; private communication with U. Ellwanger, see also
Ref.\cite{cosmology}.  For too small a value of $\mu_{\lambda}$,
{\it i.e.} very much below the typical scale $v/\sqrt{2}$,  the amount
of cold dark matter may exceed the measured value of $\Omega_{\rm CDM}
\sim 0.25$; detailed analyses are not available yet.}
compared
with the standard supersymmetry scales, $\mu=\lambda v_s/\sqrt{2}$
and/or $M_{1,2}$ for which values $\lsim O(1\, {\rm TeV})$ are
anticipated.  As a result, transparent expressions can be found by
performing a systematic expansion for small mixing between the
gauginos/doublet higgsinos and the singlino, measured by the small
size of the parameter $\mu_\lambda$ relative to the other parameters
in the mass matrix. \s

In summary, at tree--level the NMSSM neutralino sector described above
has six free parameters which we choose as $\mu_\kappa$ and
$\mu_\lambda$ in addition to the MSSM parameters: $\left\{ \{M_1, M_2,
\tan\beta, \mu\}; \mu_\kappa, \mu_\lambda\right\}$. Sometimes it is
convenient to re-express $\mu$, $\mu_\lambda$ and $\mu_\kappa$ in
terms of $\lambda$, $\kappa$ and $v_s$.  The spectrum of the NMSSM
neutralino sector will now be analyzed in detail.\s

%========================================================================
\section{NMSSM Small--Mixing Scenarios}
\label{sec:sec3}
%========================================================================

In general the diagonalization of the $5\times 5$ NMSSM mass matrix ${\cal M}_5$
cannot be performed analytically in closed form. However, if the
doublet--singlet coupling is weak, an approximate analytical solution can be
found after the $4\times 4$ MSSM submatrix ${\cal M}$ is analytically
diagonalized following the elaborate standard procedures in Ref.\cite{ckmz}.\s

The orthogonal matrix $V^5$ which transforms ${\cal M}_5$ to the
diagonal mass matrix \mbox{${\cal M}^D_5={\rm diag}\,[m_1,\ldots, m_5]$} is
conveniently split into a matrix $V$ diagonalizing the $4\times 4$
submatrix ${\cal M}$ and a matrix performing subsequently the
block diagonalization of the $4\times 4$ and $1\times 1$ submatrices. After the
block--diagonalization, the upper left MSSM mass
matrix ${\cal{M}}^D={\rm diag}\,[\tilde{m}_1, \tilde{m}_2, \tilde{m}_3,
\tilde{m}_4]$ needs
not be re-diagonalized for small doublet--singlet mixing, as
proved in the Appendix. The final result for the orthogonal matrix $V^5$
may be written in the simple form:
\begin{eqnarray}
V^5\, \approx\,
 \left(\begin{array}{cc}
    1 \hspace{-0.14cm} 1_{4\times 4}
    - \frac{1}{2}(V\Gamma) (V\Gamma)^T
  &  (V\Gamma) \\[1mm]
    -(V\Gamma)^T & 1
   -\frac{1}{2} (V\Gamma)^T (V\Gamma)
      \end{array} \right)\,
    \left(\begin{array}{cc}
          V  &  0 \\[1mm]
          0  &  1 \hspace{-0.14cm} 1_{1\times 1}
        \end{array}\right)
\label{eq:mixing_matrix}
\end{eqnarray}
The doublet--singlet 4--component mixing vector $\Gamma$ can
be expressed in terms of the gaugino/higgsino parameters as
\begin{eqnarray}
\Gamma = -\frac{\mu_\lambda}{
           {\rm det}({\cal M} - \mu_\kappa )}
          \left(\begin{array}{c}
            M'_2\, \mu\, m_Z\, s_W\, c_{2\beta}\\[1mm]
           -M'_1\, \mu\, m_Z\, c_W\, c_{2\beta}\\[1mm]
            M'_1\, M'_2 ( \mu\, c_\beta - \mu_\kappa\, s_\beta)
           -M'_{12}\, m^2_Z\, s_\beta\\[1mm]
            M'_1\, M'_2 ( \mu\, s_\beta - \mu_\kappa\, c_\beta)
                 -M'_{12}\, m^2_Z\, c_\beta
            \end{array}\right)
\label{eq:column_vector}
\end{eqnarray}
with the abbreviations
\begin{eqnarray}
  M'_1 = M_1-\mu_\kappa,\quad
  M'_2 = M_2-\mu_\kappa,\quad
  M^{(\prime)}_{12}= M^{(\prime)}_1\, c^2_W+M^{(\prime)}_2\, s^2_W
\end{eqnarray}
and the determinant
\begin{eqnarray}
{\rm det}({\cal M} - \mu_\kappa )
  =  M'_1 M'_2 (\mu^2_\kappa-\mu^2)
   + M'_{12} (\mu \,s_{2\beta}+\mu_\kappa)\, m^2_Z
\end{eqnarray}

The mixing with the singlet alters the MSSM mass eigenvalues $\tilde{m}_i$
$[i=1,\ldots 4]$ to $O(\varepsilon^2)$\footnote{Note however that small
mass differences $|\tilde{m}_i-\tilde{m}_5|$ may enhance the mixing
effects.}, and correspondingly the singlet mass
\begin{eqnarray}
\tilde{m}_5=\mu_\kappa
\end{eqnarray}
The shifts are given as
\begin{eqnarray}
&& m_i = \tilde{m}_i+ \frac{1}{\tilde{m}_i-\tilde{m}_5}
                  \,  (V X)^2_i \qquad [\,i=1-4] \nonumber\\
&& m_5 = \tilde{m}_5-\sum^4_{i}\frac{1}{\tilde{m}_i-\tilde{m}_5}
          \, (V X)^2_i
\end{eqnarray}
with the 4--component vector $X \equiv \mu_\lambda (0,0,-s_\beta, -c_\beta)$.
[The eigenvalues are not necessarily ordered
sequentially, and, if some of them are negative, the additional phase rotation
transforms them to positive physical masses.]
Even for small mixing, the 5th eigenvalue $m_5$ may differ significantly
from the singlino mass parameter $\tilde{m}_5=\mu_\kappa$ if $\kappa$ is small.
However, even though the {\it relative shift} may be large, the
{\it absolute shift} remains small, of second order.
Trivially, the eigenvalues fulfill the spur formula
\begin{eqnarray}
\sum^5_{i=1} m_i = M_1+M_2+\mu_\kappa
\end{eqnarray}
which is independent of the parameters $\mu$ and $\mu_\lambda$.\s

The doublet--singlet mixing generates a singlino component in the wave functions
of the original MSSM neutralinos $\tilde{\chi}^0_i$ $[\,i=1,\ldots,4\, ]$ of
the size
\begin{eqnarray}
V^5_{i5} \, \approx\, \sum_{j=1}^4\, V_{ij}\Gamma_j
\end{eqnarray}
linear in the mixing parameter to first approximation as expected for
off--diagonal elements. Reciprocally, the singlino component in the wave
function of $\tilde{\chi}^0_5$ is reduced to
\begin{eqnarray}
V^5_{55}\, \approx\, 1-\frac{1}{2}\,\sum_{i=1}^4 \Gamma^2_i
\end{eqnarray}
differing from unity only to second order in the mixing as expected for
diagonal elements.\s

\begin{figure}[thb]
\begin{center}
\vskip 0.3cm
\includegraphics[width=9.0cm,height=8.0cm]{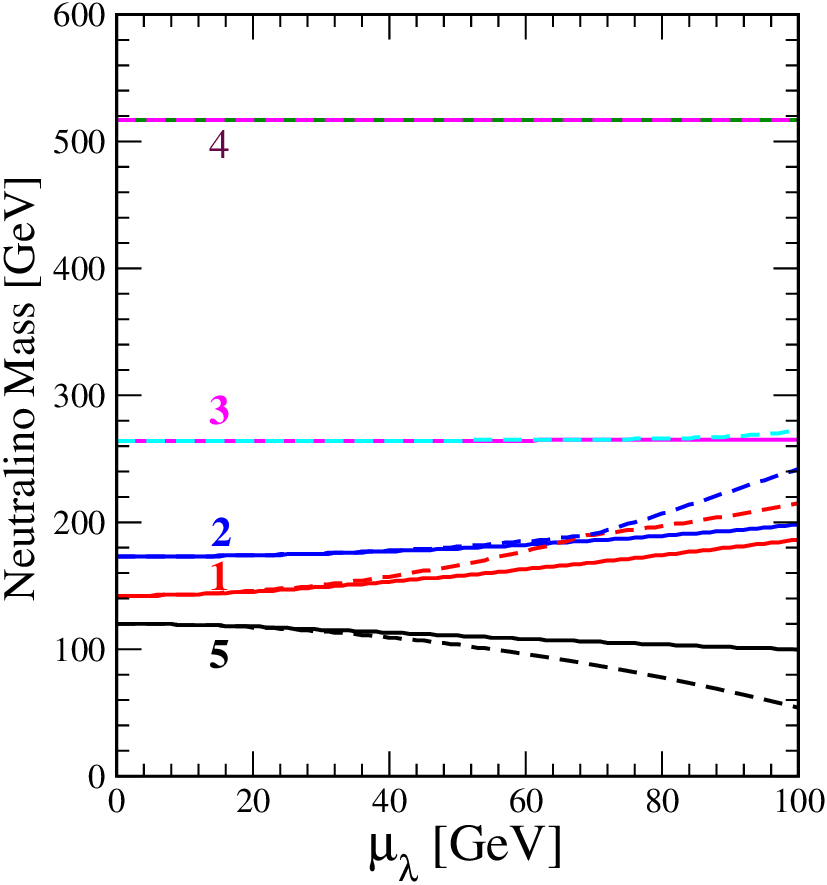}
\end{center}
\vskip -0.5cm
\caption{\it The exact numerical solution (solid) and the approximate
         solution (dashed) for the masses of the five neutralino
         states in the NMSSM as a function of $\mu_\lambda$ for the
         parameter set $\mathbb{P}=\{\mu_\kappa=120$ GeV, $M_1=250$ GeV,
         $M_2=500$ GeV, $\mu=170$ GeV, $\tan \beta=3\}$.  The
         ordering of the mass spectrum is $m_5, m_1, m_2, m_3$, and
         $m_4$ in increasing mass, {\it i.e.} the state
         $\tilde{\chi}^0_5$ is the lightest neutralino for the given
         parameter set.}
\label{fig:fig1}
\end{figure}

As long as the mixing parameter $\mu_\lambda$ is significantly smaller
than the other parameters, we find that the approximation works remarkably well,
as demonstrated in Fig.~\ref{fig:fig1}. As an example both the exact numerical
solutions and the approximate solutions for the neutralino masses
are shown as a function of $\mu_\lambda$ for a
favored parameter set $\mathbb{P}$ of broken PQ symmetry, $\mu_\kappa=120$
GeV with $M_1=250$ GeV, $M_2=500$ GeV, $\mu=170$ GeV and $\tan \beta=3$.
The exact and approximate solutions agree rather well as long as $\mu_\lambda$
is less than about 80 GeV, as the mixing corrections are of second order in
$\mu_\lambda$. \s

\begin{figure}[htb]
\begin{center}
\vskip 0.3cm
\includegraphics[width=9.0cm,height=8.0cm]{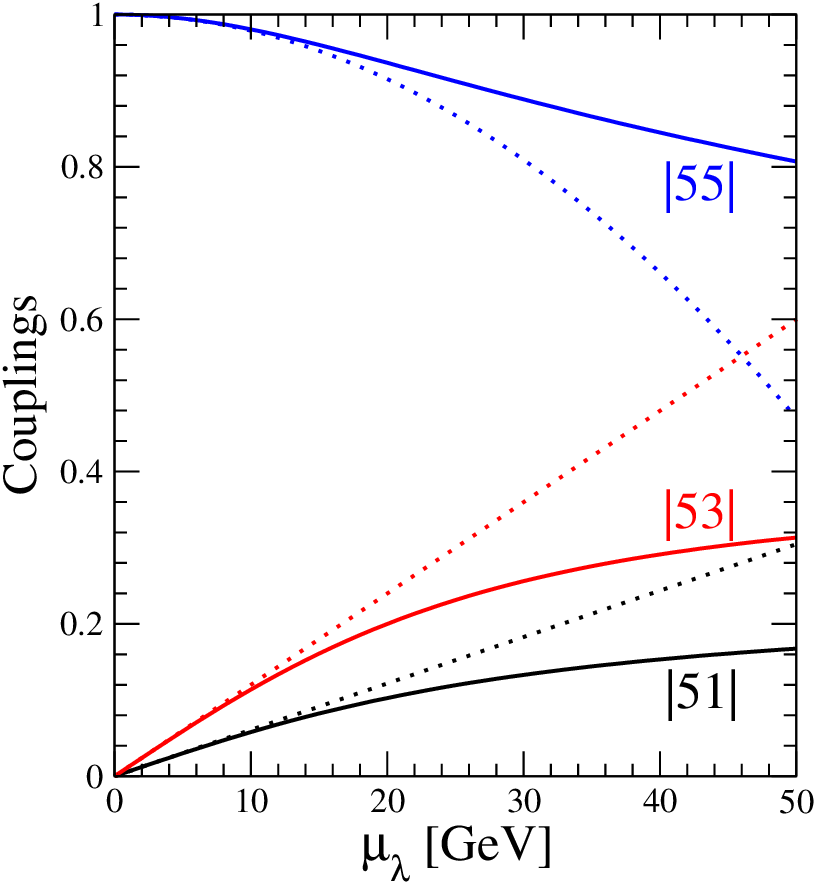}
\end{center}
\vskip -0.5cm
\caption{\it The exact numerical solution (solid) and the approximate solution
         (dashed) for the gaugino/higgsino and singlino components,
         $\{|N^5_{51}|, |N^5_{53}|, |N^5_{55}|\}$, of the lightest
         singlino--dominant neutralino as a function of $\mu_\lambda$ for
         the same parameter set $\mathbb{P}$ as in Fig.\ref{fig:fig1}.}
\label{fig:fig2}
\end{figure}

In Fig.~\ref{fig:fig2} the exact numerical solution (solid) and the
approximate solution (dashed) are compared for the gaugino/higgsino
and singlino components, $\{|(N^5)_{51}|, |(N^5)_{53}|,
|(N^5)_{55}|\}$, of the lightest singlino--dominated neutralino as a
function of $\mu_\lambda$ for the same parameter set $\mathbb{P}$.
Since the matrix $V^5$ is in general linear in the mixing term $\mu_\lambda$,
the approximate solution differs from the exact solution already for smaller
values of $\mu_\lambda$ in the reference point $\mathbb{P}$ in
which $\tilde{m}_5=\mu_\kappa$ is quite close to the higgsino parameter
$\mu$, though the characteristic features remain valid up to
$\mu_{\lambda} \sim$ 40 GeV.\s

To fully exhaust the potential of our analytical method we perform the complete
NMSSM diagonalization for the two standard limits analyzed in general within
the MSSM: $M_{1,2} \gg |\mu|$ and {\it vice versa}, both complemented of course
by small doublet--singlet mixing $\mu_\lambda \ll {\rm max}\{M_{1,2}, |\mu|\}$.\s

\subsection{Small singlino mass parameter}

The first special analysis should be performed for small singlino mass
parameter $\mu_\kappa$, which implies a slightly broken PQ symmetry
$\kappa\ll 1$ as favored by the RG flow of this coupling in grand unified
theories. Due to the small doublet--singlet mixing the structure of the original
MSSM neutralinos $\tilde{\chi}^0_i$ $[i=1 - 4]$ is changed little while the
properties of the 5th neutralino $\tilde{\chi}^0_5$, the lightest for small
$\mu_\kappa$, are determined jointly by both the singlino parameter $\mu_\kappa$
and the mixing parameter $\mu_\lambda$.\s

\subsubsection{Large gaugino mass parameters}

As a first example, we consider the case with large gaugino mass
parameters, {\it i.e.\ } \mbox{$M_{1,2}\gg |\mu| \gg m_Z,
\mu_\lambda$}. \s

To begin, the $4\times 4$ diagonalization matrix $V$ defined in
Eq.~(\ref{eq:mixing_matrix}) can be parameterized up to second
order according to standard MSSM procedure, cf. Ref.\cite{ckmz}, as
\begin{eqnarray}
V\approx
\left(\begin{array}{cc}
            V_G    &  0      \\[1mm]
              0    &  V_H
      \end{array}\right)
\left(\begin{array}{cc}
             1 \hspace{-0.14cm} 1_{2\times 2}   & V_X       \\[1mm]
              -V^T_X &   1 \hspace{-0.14cm} 1_{2\times 2}
      \end{array}\right)
\left(\begin{array}{cc}
            1 \hspace{-0.14cm} 1_{2\times 2}    & 0 \\[1mm]
            0    &  R_{\pi/4}
      \end{array}\right)
\label{eq:V_mixing}
\end{eqnarray}
The $2\times 2$ $\pi/4$ rotation
$R_{\pi/4}=(1-i\sigma_y)/\sqrt{2}$ shifts the $[34]$ off--diagonal
elements $[-\mu,-\mu]$ onto the diagonal axis $[\mu, -\mu]$.
The second matrix, $V_X$,
\begin{eqnarray}
V_X\, = \,
 \left(\begin{array}{cc}
        -c_\beta\, s_W\, m_Z/M_1  &   s_\beta\, s_W\, m_Z/M_1\\[3mm]
         c_\beta\, c_W\, m_Z/M_2  &  -s_\beta\, c_W\, m_Z/M_2
        \end{array}\right)
\end{eqnarray}
removes the mixing between the blocks of the two gaugino and the two higgsino
states. The components $V_G$ and $V_H$ diagonalize the gaugino and
higgsino blocks themselves:
\begin{eqnarray}
&& V_G \approx 1 \hspace{-0.14cm} 1_{2\times 2} -\frac{1}{2}
        \left(\begin{array}{cc}
             s^2_W \, m^2_Z/M^2_1 &  0 \\[2mm]
               0    &   c^2_W\,  m^2_Z/M^2_2
            \end{array}\right)\nonumber\\[2mm]
&& V_H \approx 1 \hspace{-0.14cm} 1_{2\times 2}-\frac{1}{2}
        \left(\begin{array}{cc}
            (1+s_{2\beta})\, M_{12}^{\prime\prime 2} m^2_Z/2 M^2_1 M^2_2
         &  0 \\[2mm]
            0
         &  (1-s_{2\beta})\, M_{12}^{\prime\prime 2} m^2_Z/2 M^2_1 M^2_2
            \end{array}\right)
\end{eqnarray}
with $M_{12}^{\prime\prime 2} = M^2_1 c^2_W+ M^2_2 s^2_W$,
respectively.  $V_G$ and $V_H$ relate to a
diagonal form of the gaugino--higgsino mass matrix for large $M_{1,2}$
and $\mu$. Their off--diagonal matrix elements are of second order and
can be omitted consistently as they would effect the eigenvalues only
to fourth order.\s

After these steps are performed, the $4\times 4$ mass submatrix is diagonal
and the complete symmetric mass matrix
${\cal M}_5$ takes the form
\begin{eqnarray}
{\cal M}_5 \rightarrow \widehat{\cal M}_5 \approx
\left(\begin{array}{cccc|c}
  \tilde{m}_1 &  &  &  & 0   \\[1mm]
   & { }\hskip 1mm \tilde{m}_2&  &  & 0   \\[1mm]
   &  & \tilde{m}_3 &  & + \mu_\lambda c_- \\[4mm]
   &  &  & \tilde{m}_4 & -\mu_\lambda c_+ \\[2mm]
   \cline{1-4}\\[-3mm]
   0 & 0 & +\mu_\lambda c_- & \multicolumn{1}{c}{-\mu_\lambda c_+ } & \mu_\kappa
      \end{array}\right)
\label{eq:1st_rotated_matrix}
\end{eqnarray}
where, in an obvious notation, zero elements are
suppressed for easier reading, and \mbox{$c_{\pm} = (c_\beta \pm s_\beta)/\sqrt{2}$}
is used as abbreviation. The MSSM neutralino mass eigenvalues are
given by
\begin{eqnarray}
&& \tilde{m}_1 = M_1 + \frac{m^2_Z}{M_1} s^2_W\nonumber\\
&& \tilde{m}_2 = M_2 + \frac{m^2_Z}{M_2} c^2_W \nonumber\\
&& \tilde{m}_3 = \mu - \frac{M_{12}}{2M_1 M_2}m^2_Z(1+s_{2\beta})
                 \nonumber\\
&& \tilde{m}_4 = -\mu - \frac{M_{12}}{2M_1 M_2}m^2_Z(1-s_{2\beta})
\end{eqnarray}
It remains to diagonalize $\widehat{\cal M}_5$ by choosing the proper form
of $V \Gamma$ in $V^5$.\s

In the limit of large gaugino mass parameters, the doublet--singlet
4--component mixing vector $\Gamma$ reduces to a simple expression
\begin{eqnarray}
\Gamma \approx \frac{\mu_\lambda}{\mu}
   \left(0,\, 0,\, c_\beta,\, s_\beta\right)^T
\label{eq:large_gaugino_gamma}
\end{eqnarray}
and the entire matrix $V^5$ can be written, up to second order, in the form
\begin{eqnarray}
V^5 \approx \left(\begin{array}{cccc|c}
  1  & { }  & { } & { } &  0 \\
  {} &   1  & { } & { } &  0 \\[2mm]
  {} &  {}  & 1- \frac{\mu^2_\lambda}{4 \mu^2} (1- s_{2\beta})  &
               { }
            & \frac{\mu_\lambda}{\mu}\, c_- \\[2mm]
  {} &  {}  &  {}
            & 1-\frac{\mu^2_\lambda}{4 \mu^2} (1+ s_{2\beta})
            & \frac{\mu_\lambda}{\mu}\,c_+
             \\[2mm]
  \cline{1-4} \\[-3mm]
   0 &  0 & -\frac{\mu_\lambda}{\mu}\,c_-
          & \multicolumn{1}{c}{-\frac{\mu_\lambda}{\mu}\,c_+ }
            & 1 - \frac{\mu^2_\lambda}{2 \mu^2}
            \end{array}\right)\,
    \left(\begin{array}{cc}
            V   &     0  \\[2mm]
            0   &     1
          \end{array}\right)
\label{eq:entire_V5}
\end{eqnarray}
with zero's suppressed in the upper $4 \times 4$ matrix, and
antisymmetric in the off--diagonal elements.\s

The rotations lead eventually to the diagonal mass matrix, of which the mass
eigenvalues to the desired order are given by
\begin{eqnarray}
&& m_1 \approx M_1 + \frac{m^2_Z}{M_1} s^2_W\nonumber\\
&& m_2 \approx M_2 + \frac{m^2_Z}{M_2} c^2_W \nonumber\\
&& m_3 \approx \mu - \frac{M_{12}}{2M_1 M_2}m^2_Z (1+s_{2\beta})
             + \frac{\mu^2_\lambda}{2\mu}(1-s_{2\beta})
         \nonumber\\
&& m_4 \approx -\mu - \frac{M_{12}}{2M_1 M_2}m^2_Z(1-s_{2\beta})
             - \frac{\mu^2_\lambda}{2\mu}(1+s_{2\beta})
         \nonumber\\
&& m_5 \approx \mu_\kappa + \frac{\mu^2_\lambda}{\mu} s_{2\beta}
\label{eq:mass_spectrum1}
\end{eqnarray}
[recall $M_{12}=M_1 c_W^2+M_2 s_W^2$]. For the ordering of the
eigenvalues and the flipping of the signs to positive physical masses
the previous general remarks apply.\s

Two points should be emphasized explicitly. While the large gaugino masses
$m_1, m_2$ are not affected by the singlino, it does affect the higgsino states
3,4 to second order. The singlino mass is also affected to second
order; however the mixing term can be leading if the singlino mass
parameter $\mu_\kappa$ is small.\s

The mixing in the wave--functions is described by the components of $\Gamma$
itself [since the $4\times 4$ matrix $V$ deviates from unity only to second
order in the small parameters of the order of the SUSY scales]:
\begin{eqnarray}
V^5_{i5} &\approx& \frac{\mu_\lambda}{\mu}\left(0,\, 0,\,
         -\frac{1}{\sqrt{2}} (c_\beta-s_\beta),\,
         -\frac{1}{\sqrt{2}} (c_\beta+s_\beta)\right)_{\!i}
                      \nonumber\\
V^5_{5i} &\approx& \frac{\mu_\lambda}{\mu}\left(0,\, 0,\, c_\beta,\,
           s_\beta\right)_{\!i}\nonumber\\
V^5_{55} &\approx& 1-\frac{\mu^2_\lambda}{2\mu^2}
\label{eq:v5}
\end{eqnarray}

\subsubsection{Large higgsino mass parameter}

As a second example, we consider the case with large higgsino mass
parameter, {\it i.e.\ } \mbox{$|\mu|\gg M_{1,2}\gg m_Z,
\mu_\lambda$}. This example is complementary to the previous case.\s

The overall diagonalization $4\times 4$ matrix $V$ can be parameterized
in the same form as that in Eq.(\ref{eq:V_mixing}).
The $2\times 2$ matrix $ V_X$ describing the mixing between the two
ensembles of the gaugino states and the higgsino states reads
\begin{eqnarray}
V_X \approx \frac{m_Z}{\mu}
 \left(\begin{array}{cc}
         s_W\,s_\beta   &  -s_W\,c_\beta \\
         -c_W\,s_\beta  &  c_W\,c_\beta
        \end{array}\right)
\end{eqnarray}
leading to a block-diagonal mass matrix composed of a $2\times 2$ matrix,
depending on $M_1$ and $M_2$ with small corrections of the order of
$m^2_Z/\mu$, and a $2\times 2$ mass matrix, depending only on the
higgsino parameter $\mu$. The $2\times 2$ blocks $V_G$ and $V_H$ in the
gaugino and higgsino sector may be written
\begin{eqnarray}
&& V_G \approx 1 \hspace{-0.14cm} 1_{2\times 2} -\frac{1}{2}
        \left(\begin{array}{cc}
            s^2_W\, m^2_Z/\mu^2
         &  0 \\[2mm]
            0
         &  c^2_W\, m^2_Z/\mu^2
            \end{array}\right) \nonumber\\[3mm]
&& V_H \approx 1 \hspace{-0.14cm} 1_{2\times 2}-\frac{1}{2}
        \left(\begin{array}{cc}
            \frac{1}{2} (1+s_{2\beta})\, m^2_Z/\mu^2
         &  0  \\[2mm]
            0
         &  \frac{1}{2} (1-s_{2\beta})\, m^2_Z/\mu^2
            \end{array}\right)
\end{eqnarray}
respectively, after the higgsino submatrix has been diagonalized by
the standard $R_{\pi/4}$ rotation.\s

These transformations diagonalize the $4\times 4$
submatrix within the block--diagonal matrix $\widehat{\cal M}_5$ of the same form
as (\ref{eq:1st_rotated_matrix}),
of which the first four diagonal elements are given by
\begin{eqnarray}
&& \tilde{m}_1 = M_1 - \frac{m^2_Z}{\mu} s^2_W\, s_{2\beta}\nonumber\\
&& \tilde{m}_2 = M_2 - \frac{m^2_Z}{\mu} c^2_W\, s_{2\beta} \nonumber\\
&& \tilde{m}_3 = \mu +\frac{m^2_Z}{2\mu} (1+s_{2\beta})
                \nonumber\\
&& \tilde{m}_4 = -\mu -\frac{m^2_Z}{2\mu} (1-s_{2\beta})
\end{eqnarray}

The mixing between the doublet--higgsino and singlino states is then
described in an analytic form by a 4--component column vector
\begin{eqnarray}
\Gamma \approx \frac{\mu_\lambda}{\mu}
   \left(0,\,  0,\,  c_\beta,\, s_\beta \right)^T
\end{eqnarray}
mixing the singlino both with the gauginos and with the doublet--higgsinos.
The entire matrix $V^5$ can be written up to second order in the form, with
antisymmetric off-diagonal elements,
\begin{eqnarray}
{ }\hskip -0.5cm V^5\! \approx
 \left(\begin{array}{cccc|c}
     1-\frac{\mu^2_\lambda m^2_Z s^2_W c^2_{2\beta}}{2M^2_1 \mu^2}
 &  { } & { } & { } & 0  \\[2mm]
   {}
 &   1-\frac{\mu^2_\lambda m^2_Z c^2_W c^2_{2\beta}}{2M^2_2 \mu^2}
 &  { } & { } & 0 \\[2mm]
   {} & {}
 & 1-\frac{\mu^2_\lambda c^2_-}{2\mu^2}
 &  { }
 &  \frac{\mu_\lambda c_-}{\mu} \\[2mm]
   {} & {} & {}
 & 1-\frac{\mu^2_\lambda c^2_+}{2 \mu^2}
 & \frac{\mu_\lambda c_+}{\mu} \\[2mm]
 \cline{1-4}
   0 & 0 & -\frac{\mu_\lambda c_-}{\mu}
         & \multicolumn{1}{c}{-\frac{\mu_\lambda c_+}{\mu} }
 & 1 - \frac{\mu^2_\lambda}{2 \mu^2}
            \end{array}\right)
    \left(\begin{array}{cc}
            V   &     0  \\[2mm]
            0   &     1
          \end{array}\right)
\end{eqnarray}
and with the same abbreviations as before in Eq.(\ref{eq:1st_rotated_matrix}).\s

The rotations lead eventually to a diagonal mass matrix,
consisting of the mass eigenvalues:
\begin{eqnarray}
&& m_1 \approx M_1 - \frac{m^2_Z}{\mu} s^2_W\, s_{2\beta}\nonumber\\
&& m_2 \approx M_2 - \frac{m^2_Z}{\mu} c^2_W\, s_{2\beta} \nonumber\\
&& m_3 \approx \mu +\frac{m^2_Z}{2\mu} (1+s_{2\beta})
                +\frac{\mu^2_\lambda}{2\mu} (1-s_{2\beta})
         \nonumber\\
&& m_4 \approx -\mu -\frac{m^2_Z}{2\mu} (1-s_{2\beta})
                -\frac{\mu^2_\lambda}{2\mu} (1+s_{2\beta})
         \nonumber\\
&& m_5 \approx \mu_\kappa + \frac{\mu^2_\lambda}{\mu} s_{2\beta}
\label{eq:mass_spectrum2}
\end{eqnarray}
with apparent reciprocity in the MSSM subsystem between gaugino and
higgsino parameters in comparison with the previous case, but
universal modifications from the doublet--singlet mixing.\s

Correspondingly, to leading order the coefficients of $V^5$ involving
the singlino index $5$ coincide with the elements of the
doublet--singlet mixing matrix in Eq.(\ref{eq:v5}).

\subsection{Large singlino mass parameter}

In the alternative extreme, the PQ symmetry is strongly broken if
$\kappa$ is large and, equivalently, the singlino mass parameter is
large, i.e. $\mu_\kappa \gg \mu_\lambda, \mu, M_{1,2}$.
This limit is not favored by the renormalization group flow from the
GUT scale down to the electroweak scale but
cannot be ruled out {\it a priori} on general grounds. The new fifth
eigenstate, predominantly composed of the singlino, would in
general be the heaviest state, mixed only weakly with the
iso--doublets and, as a result, coupling weakly to electroweak
gauge bosons and matter fields.\s

Applying the approximation method described in the Appendix and the
general introduction to this section, the neutralino mass matrix can
be transformed into the $4\times 4$ and $1\times 1$ block--diagonal
form by inserting the mixing column vector
\begin{eqnarray}
\Gamma \approx \frac{\mu_\lambda}{\mu_\kappa}\,
                    \left(\,0,\, 0,\, s_\beta,\, c_\beta\,\right)^T
\label{eq:mixing_vector_Ls}
\end{eqnarray}
in the $V^5$ matrix Eq.(\ref{eq:mixing_matrix}). Note that the mixing column
vector (\ref{eq:mixing_vector_Ls}) is directly proportional to the 4--component
off--diagonal column vector of the mass matrix (\ref{eq:mass_matrix})
unlike the column vector (\ref{eq:large_gaugino_gamma}) for a small singlino
mass parameter.\s

From the general analysis it is apparent that the first four neutralino
masses, of MSSM type, are modified to the order $\mu^2_\lambda/\mu_\kappa$
through the higgsino part, as is the 5th neutralino mass. The mass and
the $55$ wave--function are approximately given by
\begin{eqnarray}
m_5\approx \mu_\kappa + \frac{\mu^2_\lambda}{\mu_\kappa}
\end{eqnarray}
and
\begin{eqnarray}
V^5_{55}\,\approx\,1-\frac{\mu^2_\lambda}{2\,\mu^2_\kappa}
\end{eqnarray}
while doublet components are mixed in to first order,
\begin{eqnarray}
V^5_{5i} \approx - \frac{\mu_\lambda}{\mu_\kappa}\,
                    \left(\,0,\, 0,\, s_\beta,\, c_\beta\,\right)
\end{eqnarray}
in parallel to the singlino components of the first doublet--type
neutralinos.\s

In summary, the gaugino/doublet higgsino dominated neutralinos follow the
pattern of the MSSM quite narrowly. Increasing the value of $\mu_\kappa$ will
increase the mass of the new singlino state (almost) linearly, causing the
state to decouple and making the NMSSM very difficult to distinguish
from the MSSM.\s

\subsection{The case with $M_1=M_2$ in the limit of $\tan\beta=1$}

When the two soft--breaking SU(2) and U(1) gaugino masses are
equal, $M_1=M_2=M$ and $\tan\beta=1$, cf. Ref.\cite{ckmz}, the electroweak
gauge symmetry guarantees the existence of a physical neutral
state which does not mix with the other states and which has a
mass eigenvalue identical to the modulus $M$. Furthermore, the
gaugino states do not mix with the singlino state $\widetilde{S}$
that couples only to the specific linear combination of the
higgsino states $\widetilde{H}^0_b =(\widetilde{H}^0_u
+\widetilde{H}^0_d)/\sqrt{2}$. As a result, one gaugino state
mixes only with one higgsino state while the other orthogonal
higgsino state mixes with the singlino state, leading to a
block-diagonal matrix composed of one scalar and two $2\times 2$
matrices.\s

This special structure can be made apparent by switching to the mixed
basis $\{\tilde{\gamma}, \tilde{Z}, \tilde{H}^0_a, \tilde{H}^0_b,
\tilde{S}\}$ from the original group basis $\{\tilde{B}, \tilde{W}^3,
\tilde{H}^0_d, \tilde{H}^0_u, \tilde{S}\}$ by means of the
transformation
\begin{eqnarray}
\left(\begin{array}{c}
     \tilde{\gamma} \\
     \tilde{Z}      \\
     \tilde{H}_a    \\
     \tilde{H}_b    \\
     \tilde{S}
      \end{array}\right)
={\cal A}_5
\left(\begin{array}{l}
    \tilde{B}   \\
    \tilde{W}^3 \\
    \tilde{H}_d \\
    \tilde{H}_u \\
    \tilde{S}
     \end{array}\right)
= \left(\begin{array}{rccrc}
    c_W  & s_W & 0  &  0 & 0 \\
    -s_W & c_W & 0  &  0 & 0 \\
     0   &  0  & \frac{1}{\sqrt{2}}
               & -\frac{1}{\sqrt{2}} & 0 \\
     0   &  0  & \frac{1}{\sqrt{2}}
               & \frac{1}{\sqrt{2}}  & 0 \\
     0   &  0  &   0  & 0  & 1
        \end{array}\right)
\left(\begin{array}{l}
    \tilde{B}   \\
    \tilde{W}^3 \\
    \tilde{H}_d \\
    \tilde{H}_u \\
    \tilde{S}
     \end{array}\right)
\end{eqnarray}
In this new $\{\tilde{\gamma}, \tilde{Z}, \tilde{H}^0_a,
\tilde{H}^0_b, \tilde{S}\}$ basis the mass matrix $\widehat{\cal
M}_5$ takes the block-diagonal form
\begin{eqnarray}
\widehat{\cal M}_5 = {\cal A}_5 {\cal M}_5 {\cal A}_5^T
  = \left(\, \begin{array} {ccccc}\cline{1-1}
  \multicolumn{1}{|c|}{M}
               &   0   &  0   &   0   &  0  \\
          \cline{1-3}
  \multicolumn{1}{c|}{0}  &   M   &
  \multicolumn{1}{c|}{m_Z} &   0   &  0  \\
  \multicolumn{1}{c|}{0}   &  { }\, m_Z &
  \multicolumn{1}{c|}{\mu} &   0   &  0  \\
          \cline{2-5}
           0    &   0   &   0  &
  \multicolumn{1}{|c}{-\mu}  &
  \multicolumn{1}{c|}{-\mu_\lambda} \\
           0    &   0   &   0  &
  \multicolumn{1}{|c}{{ }\, -\mu_\lambda} &
  \multicolumn{1}{c|}{\mu_\kappa}\\
         \cline{4-5}
           \end{array}\, \right)
\end{eqnarray}
This mass matrix generates two two--state mixings between
$\tilde{Z}$ and $\tilde{H}^0_a$, and between $\tilde{H}^0_b$ and
$\tilde{S}$, respectively. The block--diagonal matrix can be
diagonalized by the orthogonal matrix
\begin{eqnarray}
\hat{V}^5 =\left(\begin{array}{ccc}
     1   &   { }   &   { } \\
     { } & R_{g/h} &   { } \\
     { } &   { }   & R_{h/s}
     \end{array}\right)
\end{eqnarray}
consisting of two $2\times 2$ rotation matrices
$R_{g/h}$ and $R_{h/s}$,
\begin{eqnarray}
R_{g/h} = \left(\begin{array}{cr}
           \cos\theta_{g/h}  & -\sin\theta_{g/h} \\
           \sin\theta_{g/h}  &  \cos\theta_{g/h}
                \end{array}\right),\quad
R_{h/s} = \left(\begin{array}{cr}
           \cos\theta_{h/s}  & -\sin\theta_{h/s} \\
           \sin\theta_{h/s}  &  \cos\theta_{h/s}
                \end{array}\right)
\end{eqnarray}
with the mixing angles determined by the relations
\begin{eqnarray}
&& \tan\theta_{g/h} = -\frac{2\, m_Z}{M-\mu-\sqrt{(M-\mu)^2+4m^2_Z}}\nonumber\\
&& \tan\theta_{h/s} = \frac{2\, \mu_\lambda}{\mu_\kappa+\mu
                    +\sqrt{(\mu_\kappa+\mu)^2+4 \mu^2_\lambda}}
\end{eqnarray}

The mass eigenvalues can be written completely in analytic form,
\begin{eqnarray}
&& m_1 = M\nonumber\\
&& m_2 = \frac{1}{2}\left(M+\mu+\sqrt{(M-\mu)^2+4 m^2_Z}\right)\nonumber\\
&& m_3 = \frac{1}{2}\left(M+\mu-\sqrt{(M-\mu)^2+4 m^2_Z}\right)\nonumber\\
&& m_4 = \frac{1}{2}\left(\mu_\kappa-\mu-\sqrt{(\mu_\kappa+\mu)^2
          + 4 \mu^2_\lambda} \right)\nonumber\\
&& m_5 = \frac{1}{2}\left(\mu_\kappa-\mu+\sqrt{(\mu_\kappa+\mu)^2
          + 4 \mu^2_\lambda} \right)
\end{eqnarray}
with the wave--functions of the neutralinos $\tilde{\chi}^0_i$ $[ i=1,\ldots,
5]$ determined by the cos/sin of the mixing angles $\theta_{g/h}$ and
$\theta_{h/s}$. \s

\begin{figure}[ht]
\begin{center}
\vskip 0.3cm
\includegraphics[width=14.0cm,height=8.0cm]{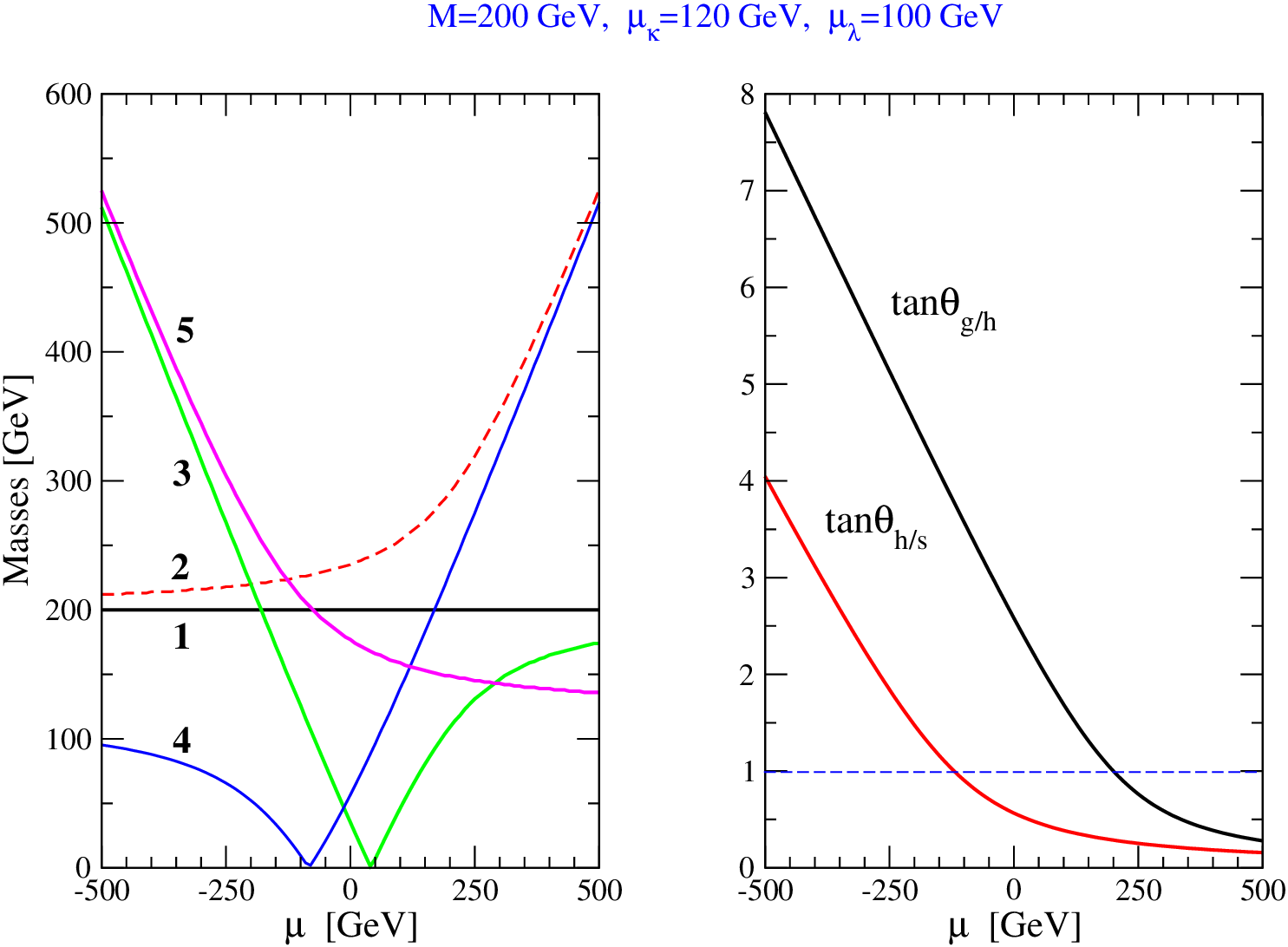}
\end{center}
\vskip -0.5cm \caption{\it (a) The neutralino masses $|m_i|$, mapped onto
         positive values, and (b) the tangent values of
         the mixing angles $\theta_{g/h}$ and $\theta_{h/s}$ as a function of
         the higgsino mass parameter $\mu$ for the parameter set:
         $M=200$ GeV, $\mu_\kappa=120$ GeV, $\mu_\lambda=100$ GeV.}
\label{fig:cpmapp3}
\end{figure}

It is instructive to study the neutralino mass spectrum in this model for
a set of fixed parameters $M=200$ GeV and $\mu_\kappa=120$ GeV,
$\mu_\lambda=100$ GeV, by varying the higgsino mass parameter $\mu$.
The branch character of the eigenvalues $\{23\}$ and $\{45\}$
is exemplified in
Fig.\ref{fig:cpmapp3}(a). The tan's of the corresponding mixing angles are
displayed in Fig.\ref{fig:cpmapp3}(b). With rising tan's we move from scenarios
of no mixing, to maximal gaugino/doublet higgsino in $\{23\}$ and
doublet/singlet higgsino mixing in $\{45\}$, finally to gaugino/doublet
higgsino flipping $\{23\}$, and doublet/singlet flipping $\{45\}$, while
the gaugino $\{1\}$ remains untouched.\s

%===========================================
\section{Neutralino Production and Decays}
\label{sec:sec4}
%===========================================

In the MSSM the neutralino sector consists of two gauginos and two higgsinos.
Typically the lightest supersymmetric particle (LSP), which is stable under
the assumption of $R$-parity conservation, is the lightest state
of the neutralino mass matrix. The LSP will appear as one of the final states
of each sparticle decay and its non--observability is responsible
for the well--known missing energy/momentum signature of supersymmetric particle
production.  \s

The neutralino production and decay properties in the NMSSM with the additional
singlino state depend crucially on the singlino mass with respect to
the MSSM neutralino masses \cite{nmssm2}. If the singlino is much heavier than
the other states, it will be very rarely produced and so practically
unobservable. On the contrary, if the singlino is lighter than the other states,
a singlino--dominated state will be the LSP so that the other neutralino
states will decay, possibly through cascades, into the singlino--dominated
LSP. \s

In this section, we present a qualitative description of the production of
neutralinos, involving at least one singlino--dominated state, such as
$\tilde{\chi}^0_5\tilde{\chi}^0_5$ and $\tilde{\chi}^0_1\tilde{\chi}^0_5$
and the subsequent decays of the neutralino $\tilde{\chi}^0_1$ into
leptons and light Higgs bosons.\s

%============================================================
\subsection{Singlino Production in $e^+e^-$ Annihilation}
\label{sec:sec4.1}
%============================================================

The production processes
\begin{eqnarray}
e^+e^-\rightarrow \tilde{\chi}^0_i\tilde{\chi}^0_j\quad [i,j=1-5]
\end{eqnarray}
are generated by $s$--channel $Z$ exchange, and $t$-- and $u$--channel
$\tilde{e}_{L,R}$ exchanges.  After appropriate Fierz transformations
of the selectron exchange amplitudes [with the electron mass
neglected], the transition matrix element of the production process
can be written as
\begin{equation}
T(e^+e^-\rightarrow\tilde{\chi}^0_i\tilde{\chi}^0_j) =
\sum_{\alpha,\beta\,=L,R} Q_{\alpha\beta}
  \left[\bar{v}(e^+)\gamma_\mu u(e^-)\right]_{\alpha}\,
  \left[\bar{u}(\tilde{\chi}^0_i)\gamma^\mu v(\tilde{\chi}^0_j)\right]_{\beta}
\label{eq:amplitude1}
\end{equation}
The transition amplitudes are built up by the sum of the products of
chiral neutralino currents and chiral fermion currents.
The four generalized bilinear charges
$Q_{\alpha\beta}$ correspond to independent helicity amplitudes, describing
the neutralino production processes for polarized
electrons/positrons \cite{ckmz}.
They are defined by the fermion and neutralino currents and the propagators
of the exchanged (s)particles as follows:
\begin{eqnarray}
&& Q_{LL}=+\frac{D_Z}{s_W^2c_W^2}\,
           (I^f_3 - Q_f s_W^2) {\cal Z}_{ij}
            -D_{uL}g_{Lij}, \quad
 Q_{RL}=-\frac{D_Z}{c_W^2}\,
            Q_f {\cal Z}_{ij}
            +D_{tR}g_{Rij}\nonumber\\
&& Q_{LR}=-\frac{D_Z}{s_W^2c_W^2}\,
           (I^f_3-Q_f s^2_W){\cal Z}^*_{ij}
            +D_{tL}g^*_{Lij},\quad
 Q_{RR}=+\frac{D_Z}{c_W^2} Q_f {\cal Z}^*_{ij}
            -D_{uR}g^*_{Rij}
\end{eqnarray}
with $f=e^-$ in the production channel.
The first term in each bilinear charge is
generated by $Z$ exchange and the second term by selectron
exchange; $D_Z$, $D_{tL,R}$ and $D_{uL,R}$ denote the $s$--channel $Z$
propagator and the $t$-- and $u$--channel left/right--type selectron
propagators
\begin{eqnarray}
 D_Z=\frac{s}{s-m^2_Z+im_Z\Gamma_Z},\qquad
 D_{(t,u)L,R}=\frac{s}{(t,u)-m^2_{\tilde{f}_{L,R}}}
\end{eqnarray}
with $s=(p_{e^-}+p_{e^+})^2$, $t=(p_{e^-}-p_{\tilde{\chi}^0_i})^2$ and
$u=(p_{e^-}-p_{\tilde{\chi}^0_j})^2$ representing the Mandelstam variables
for neutralino pair production in $e^+e^-$ collisions.
Finally, the matrices ${\cal Z}_{ij}$,
$g_{Lij}$ and $g_{Rij}$ are products of the neutralino diagonalization
matrix elements $N^5_{ij}$
\begin{eqnarray}
&& {\cal Z}_{ij}=(N^5_{i3}N^{5*}_{j3}-N^5_{i4}N^{5*}_{j4})/2\nonumber\\
&& g_{Lij}=\left(I^f_3 N^5_{i2} c_W+(Q_f-I^f_3) N^5_{i1}s_W\right)
           \left(I^3_f N^5_{j2}c_W+(Q_f-I^f_3) N^5_{j1}s_W\right)^*/
                s_W^2c_W^2\nonumber\\
&& g_{Rij}=Q^2_f N^5_{i1}N^{5*}_{j1}/c_W^2
\label{eq:Zgg}
\end{eqnarray}

\begin{figure}[ht]
\begin{center}
\vskip 0.3cm
\includegraphics[height=8.0cm]{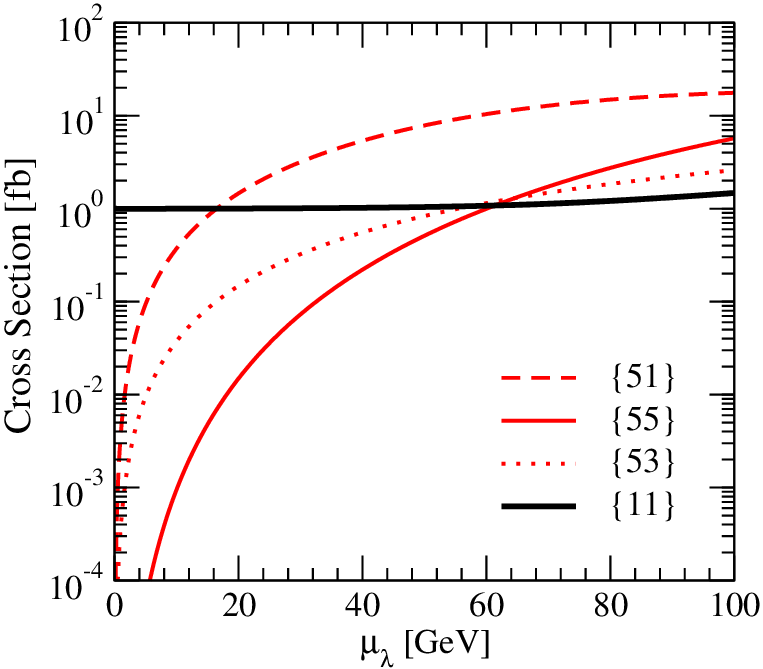}
\end{center}
\vskip -0.5cm
\caption{\it The production cross sections of neutralino pairs,
         $\{51\}$ (dashed), $\{55\}$ (thin--solid), $\{53\}$ (dotted)
         and $\{11\}$ (thick--solid), in $e^+e^-$ collisions with the
         center-of-mass energy $\sqrt{s}=500$ GeV as a function of
         $\mu_\lambda$ for the parameter set $\mathbb{P}$
         [Fig.\ref{fig:fig1}] with \mbox{$m_{\tilde{e}_R}=200$~GeV} and
         $m_{\tilde{e}_L}=250$~GeV.}
\label{fig:cpmapp4}
\end{figure}

The $e^+e^-$ annihilation cross sections follow from the squares of the bilinear
charges,
\begin{eqnarray}
\sigma\left[e^+e^-\rightarrow \tilde{\chi}^0_i \tilde{\chi}^0_j\right]
 &=&{\cal S}_{ij}\, \frac{\pi \alpha^2}{2 s}\lambda_{\rm PS}^{1/2}
  \int^1_{-1}
      \bigg\{\left[ 1-(\mu^2_i-\mu^2_j)^2
           +\lambda_{\rm PS} \cos^2\Theta\right] Q_1
     \nonumber\\
  && \hskip 3.cm   + 4\mu_i\mu_j Q_2
           +2\lambda_{\rm PS}^{1/2} Q_3 \cos\Theta\bigg\}
    \, d\cos\Theta\,
\end{eqnarray}
where ${\cal S}_{ij}$ is a statistical factor: 1 for $i\neq j$ and
$1/2$ for $i=j$; $\mu_i= m_{\tilde{\chi}^0_i}/\sqrt{s}$, $\Theta$ is
the polar angle of the produced neutrinos; and $\lambda_{\rm PS} =
\lambda_{\rm PS}(1,\mu^2_i,\mu^2_j)$ denotes the familiar $2$--body phase space
function $\lambda_{\rm PS}(x,y,z) \equiv x^2+y^2+z^2-2xy-2xz-2yz$. The quartic
charges $Q_i$ ($i=1,2,3$) are given by the bilinear charges as follows:
\begin{eqnarray}
&& Q_1 = \frac{1}{4}\left[ |Q_{RR}|^2+|Q_{LL}|^2
                           +|Q_{RL}|^2+|Q_{LR}|^2\right]\nonumber\\
&& Q_2 =\frac{1}{2} \real \left[ Q_{RR} Q^*_{RL} + Q_{LL} Q^*_{LR}
                          \right]\nonumber\\
&& Q_3 = \frac{1}{4}\left[ |Q_{RR}|^2+|Q_{LL}|^2
                           -|Q_{RL}|^2-|Q_{LR}|^2\right]
\label{eq:quartic_charges}
\end{eqnarray}
An example for the production of $\tilde{\chi}^0_5$ in association
with another singlino--type ($\tilde{\chi}^0_5$), or a gaugino--type
($\tilde{\chi}^0_1$) or a higgsino--type ($\tilde{\chi}^0_3$)
neutralino is presented in Fig.\ref{fig:cpmapp4} for the parameter
set $\mathbb{P}$ [Fig.\ref{fig:fig1}] with $m_{\tilde{e}_R}=200$ GeV
and $m_{\tilde{e}_L}=250$ GeV.  [Of course the $\{55\}$ final state is
unobservable without additional ISR $\gamma$ emission.]  The increase
of the cross sections with increasing doublet--singlet
gaugino/higgsino mixing parameterized by $\mu_\lambda$ is obvious. The
gaugino character of $\tilde \chi_1^0$ is responsible for the dominant
size of the $\{51\}$ cross-section. \s

With the anticipated integrated luminosity $\int {\cal L} = 1$ ab$^{-1}$,
sufficiently large event rates of order 10$^3$ are predicted if $\mu_\lambda$
is not too small.\s

%============================================================
\subsection{Decays to a Singlino, with no Higgs bosons}
\label{sec:sec4.2}
%============================================================

{\bf (i)} If kinematically allowed, two--body decays of neutralinos to the
electroweak gauge bosons $Z$ are among the dominant channels. The widths of
decays $\tilde{\chi}^0_i \rightarrow \tilde{\chi}^0_j Z$ are given by
\begin{eqnarray}
\Gamma[\tilde{\chi}^0_i\!\rightarrow\!\tilde{\chi}^0_j Z] =
\frac{g^2 \lambda_{\rm PS}^{1/2}}{16\pi m_{\tilde{\chi}^0_i}}
   \left\{ |{\cal Z}^2_{ij}|\!\!
         \left[\frac{(m^2_{\tilde{\chi}^0_i}\!-\!m^2_{\tilde{\chi}^0_j})^2}{
                    m^2_Z}\!+\!m^2_{\tilde{\chi}^0_i}\!+\!m^2_{\tilde{\chi}^0_j}
                    \!-\!2 m^2_Z\right]
         \!+\! 6 m_{\tilde{\chi}^0_i} m_{\tilde{\chi}^0_j}
           \real({\cal Z}^2_{ij})\!\right\}
\end{eqnarray}
where $\lambda_{\rm PS} =\lambda_{\rm PS}(1,m^2_{\tilde{\chi}^0_j}/
m^2_{\tilde{\chi}^0_i}, m^2_Z/m^2_{\tilde{\chi}^0_i})$, with ${\cal Z}_{ij}$
defined in Eq.(\ref{eq:Zgg}). The widths of the chargino 2-body decays into
a neutralino and a $W$ boson, $\tilde{\chi}^\pm_i\rightarrow
\tilde{\chi}^0_j\, W^\pm$, read correspondingly
\begin{eqnarray}
&& \hskip -0.8cm\Gamma[\tilde{\chi}^\pm_i\rightarrow \tilde{\chi}^0_j\, W^\pm] =
\frac{g^2 \lambda_{\rm PS}^{1/2}}{16\pi m_{\tilde{\chi}^\pm_i}}
   \left\{\frac{|{\cal W}_{Lij}|^2+|{\cal W}_{Rij}|^2}{2}
%                                    \overline{|{\cal W}_{ij}|^{2}}
         \left[\frac{(m^2_{\tilde{\chi}^\pm_i}-m^2_{\tilde{\chi}^0_j})}{
                    m^2_W}+m^2_{\tilde{\chi}^\pm_i}+m^2_{\tilde{\chi}^0_j}
                    -2 m^2_W\right]\right.\nonumber\\
       &&\left.\hskip 5.cm  -6\, m_{\tilde{\chi}^\pm_i} m_{\tilde{\chi}^0_j}
           \,\real({\cal W}_{Lij}{\cal W}^*_{Rij})\right\}
\end{eqnarray}
where $\lambda_{\rm PS}=\lambda_{\rm PS}(1, m^2_{\tilde{\chi}^0_j}/
m^2_{\tilde{\chi}^\pm_i}, m^2_W/m^2_{\tilde{\chi}^\pm_i})$ and the bilinear
charges ${\cal W}_{L,R}$ are defined as
\begin{eqnarray}
&& {\cal W}_{Lij}= U^*_{Li1} N^5_{j2}
                 +\frac{1}{\sqrt{2}} U^*_{Li2} N^5_{j3},\quad
   {\cal W}_{Rij}= U^*_{Ri1} N^{5*}_{j2}
                 -\frac{1}{\sqrt{2}} U^*_{Ri2} N^{5*}_{j4}
%\\
%&& \overline{|{\cal W}_{ij}|^{2}}
%                 =\left(|{\cal W}_{Lij}|^2+|{\cal W}_{Rij}|^2\righ t)/2
\end{eqnarray}
The unitary matrices $U_L$ and $U_R$ diagonalize the chargino mass
matrix as \mbox{$U_R {\cal M}_C U^\dagger_L ={\rm diag}
\left\{m_{\tilde{\chi}^\pm_1}, m_{\tilde{\chi}^\pm_2}\right\}$,} cf.
Ref.\cite{ref:chargino} for details.\s

If 2--body decay channels are closed kinematically, the 3--body neutralino
decays, \mbox{$\tilde{\chi}^0_i\rightarrow\tilde{\chi}^0_j f\bar{f}$}, are generated
by $s$--channel (virtual) $Z$ exchange, and $t$-- and $u$--channel sfermion
exchanges. Neglecting fermion masses, the transition matrix element, cf.
Ref.\cite{ref:CSS}, is determined by the bilinear charges
$Q^\prime_{\alpha\beta}$ which are related to the bilinear charges
$Q_{\alpha\beta}$ introduced for the production, by crossing symmetry as
\begin{eqnarray}
Q'_{\alpha\beta} = Q^*_{\alpha\beta}
\end{eqnarray}
with the transformed Mandelstam variables, $s=(p_f+p_{\bar{f}})^2$,
$t=(p_{\tilde{\chi}^0_j}+p_{\bar{f}})^2$ and
\mbox{$u=(p_{\tilde{\chi}^0_j}+p_f)^2$} for the decays. [Neutralino decays to
charginos and $W$ bosons can be described in the same way after obvious
redefinitions of the bilinear charges.] Decay widths and distributions
depend on the quartic charges $Q^\prime_1$, $Q^\prime_2$ and $Q^\prime_3$
defined analogously to Eq.(\ref{eq:quartic_charges}).\s

{\bf (ii)} At the LHC, cascade sfermion decays, $\tilde{f}\rightarrow f
\tilde{\chi}^0_i$, are of great experimental interest. The width of the
sfermion 2-body decay into a fermion and a neutralino follows from
\begin{eqnarray}
\Gamma[\tilde{f}\rightarrow f \tilde{\chi}^0_i]=
\frac{g^2 \lambda_{\rm PS}^{1/2}}{16\pi m_{\tilde{f}}} \, |g_{\tilde{f}i}|^2
\left(m^2_{\tilde{f}}-m^2_{\tilde{\chi}^0_i}-m^2_f\right)
\end{eqnarray}
where the 2--phase space function $\lambda_{\rm PS} = \lambda_{\rm PS}(1,
m^2_{\tilde{\chi}^0_i}/m^2_{\tilde{f}},
m^2_f/m^2_{\tilde{f}})$ with $\tilde{f}=\tilde{f}_L$ or $\tilde{f}_R$;
the couplings are expressed in terms of the neutralino mixing matrix $N^5$ as
\begin{eqnarray}
g_{\tilde{f}_L i} = \sqrt{2} \left[ I^f_3 N^5_{i2}  +
                   (Q_f-I^f_3) N^5_{i1} t_W\right]\quad {\rm and}\quad
g_{\tilde{f}_R i} = 2 Q_f s_W N^{5*}_{i1}
\end{eqnarray}
in obvious notation.\s

The reverse decays, neutralino [chargino] decays to sfermions plus fermions,
$\tilde{\chi}^0_i\rightarrow\tilde{f}f$ etc, are given by the corresponding
partial widths,
\begin{eqnarray}
\Gamma[\tilde{\chi}^0_i\rightarrow\tilde{f}f]
=\frac{g^2 \lambda_{\rm PS}^{1/2}}{32\pi\, m_{\tilde{\chi}^0_i}} \,
 |g_{\tilde{f}i}|^2
  \left(m^2_{\tilde{\chi}^0_i}+m^2_f-m^2_{\tilde{f}}\right)
\end{eqnarray}
with the same couplings as before and $\lambda_{\rm PS}= \lambda_{\rm PS}(1,
m^2_{\tilde{f}}/m^2_{\tilde{\chi}^0_i}, m^2_f/m^2_{\tilde{\chi}^0_i})$.
[Analogous expressions hold for chargino decays.]
\s

\begin{figure}[ht!]
\begin{center}
\vskip 0.3cm
\includegraphics[height=13.0cm]{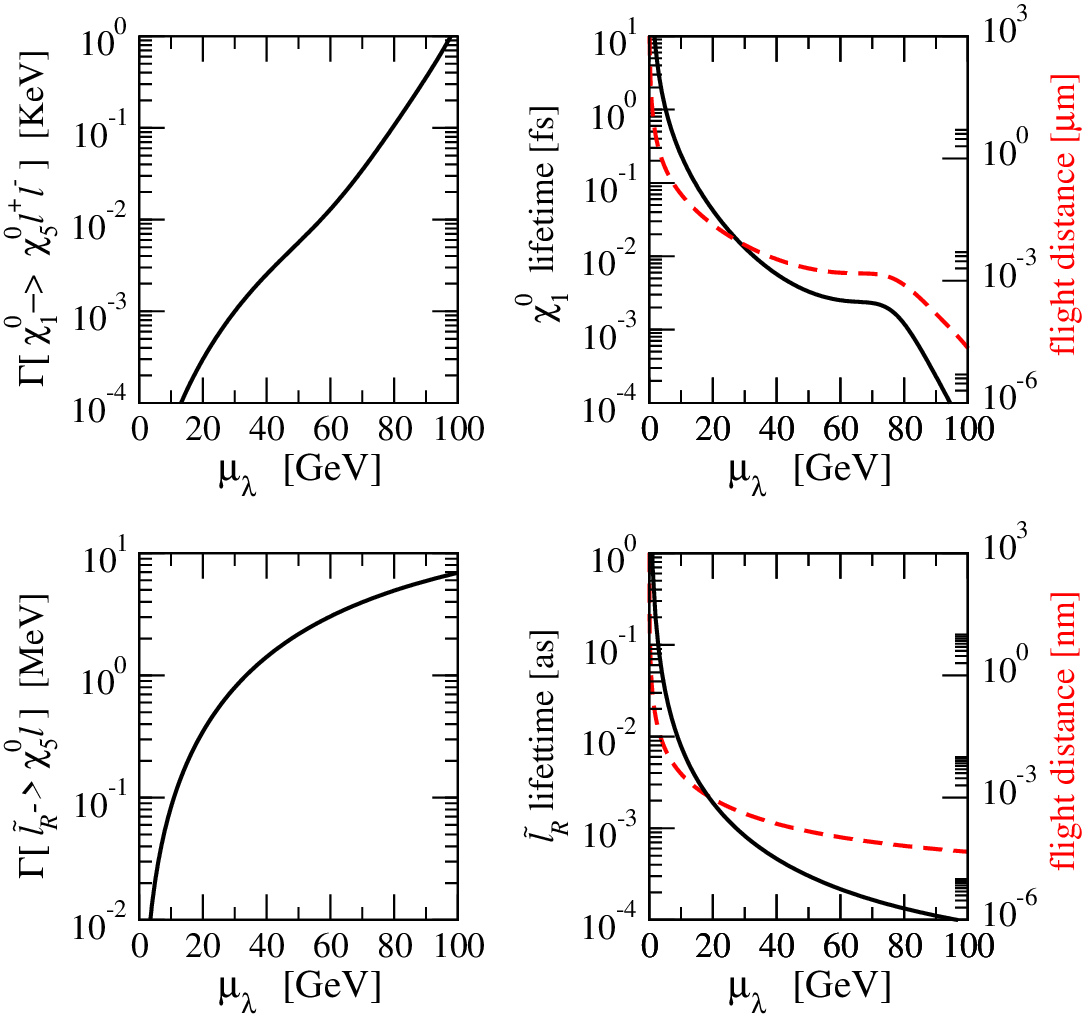}
\end{center}
\vskip -0.5cm
\caption{\it The widths, lifetimes and flight distances [broken lines]
             of the decays $\tilde{\chi}^0_1\rightarrow
             \tilde{\chi}^0_5\, l^+l^-$ and $\tilde{l}_R\rightarrow
             \tilde{\chi}^0_5 l$ as a function of $\mu_\lambda$ for
             the parameter set $\mathbb{P}$[Fig.\ref{fig:fig1}]. The
             mass of the right--handed slepton is taken to be
             $m_{\tilde{l}_R}=200 \, {\rm GeV} > m_{\tilde \chi_1^0}$
             for the 3--body neutralino decays (upper panels) and to
             be $m_{\tilde{l}_R}=130 \, {\rm GeV} < m_{\tilde
             \chi_1^0}$ for the 2--body slepton decays (lower
             panels). The masses of the squarks are
             assumed to be $m_{\tilde{q}_L}=250$ GeV and
             $m_{\tilde{q}_R}=200$ GeV.  Right: flight distances for
             $\sqrt{s}=500$ GeV are shown by broken lines.
             The kink in the $\tilde \chi_1^0$ lifetime
             and flight distance in the upper right panel
             is caused by accidental cancellations between
             sfermion and $Z$ exchange diagrams. [The value
             of the lower bound, expected from cosmological arguments
             on $\mu_\lambda$ is presently not yet known.]}
\label{fig:cpmapp5}
\end{figure}

Examples of these partial decay widths are shown in
Fig.\ref{fig:cpmapp5} [with the parameter set $\mathbb{P}$ as in
Fig.\ref{fig:fig1}]. For an illustrative purpose, the mass of the
R--sleptons $\tilde{l}_R$ is assumed to be $m_{\tilde{l}_R}=200 \,
{\rm GeV} > m_{\tilde \chi_1^0}$ for the 3--body neutralino decays and
to be $m_{\tilde{l}_R}=130 \, {\rm GeV} < m_{\tilde \chi_1^0}$ for the
2--body slepton decays, respectively.\footnote{In either case
$\tilde{\chi}^0_1$ or $\tilde{l}_R$ is the next--to--lightest SUSY
particle NLSP with just one decay channel open to the lightest SUSY
particle LSP$=\tilde{\chi}^0_5$.}  The masses of the
squarks are assumed to be $m_{\tilde{q}_L}=250$ GeV and
$m_{\tilde{q}_R}=200$ GeV.  For small mixing $\mu_{\lambda}$ the
lifetimes of the second lightest neutralino $\tilde \chi^0_1$ and the
R-sleptons $\tilde l_R$ can be quite large, giving rise potentially to
macroscopic flight paths \cite{ref:EH}.  However, cosmological bounds
on $\mu_\lambda$ must be analyzed before any (realistic) experimental
conclusions can be drawn. The kink in the $\tilde \chi_1^0$ lifetime
and flight distance in the upper right panel of
Fig.\ref{fig:cpmapp5} is caused by accidental cancellations between
sfermion and $Z$ exchange diagrams in the decays $\tilde \chi_1^0 \to
\tilde \chi_5^0 \, q \bar q$ and $\tilde \chi_5^0 \, \nu \bar \nu$;
these accidental cancellations do not occur [to any significant
degree] in the decay $\tilde \chi_1^0 \to \tilde \chi_5^0 \, l^+l^-$.
%

%============================================================
\subsection{Decays to a Singlino, involving Higgs bosons}
\label{sec:sec4.3}
%============================================================

\begin{figure}[t]
\begin{center}
\vskip 0.5cm
\includegraphics[height=8.0cm,clip=true]{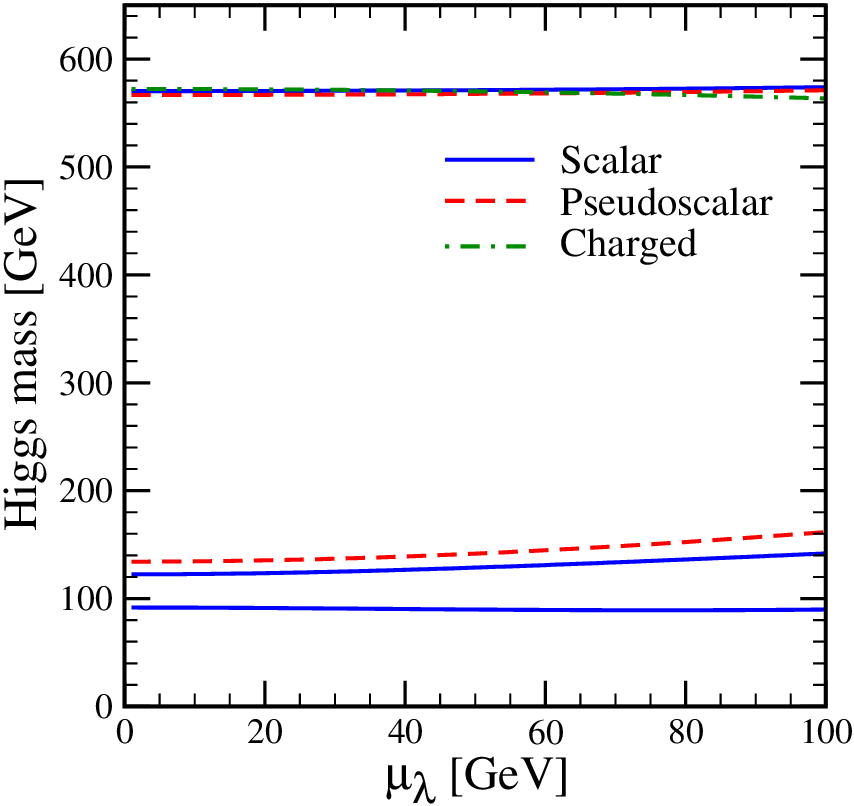}
\end{center}
\vskip -0.5cm
\caption{\it The Higgs boson mass spectrum as a function of
         $\mu_{\lambda}$ for the parameter set
         $\mathbb{P}$[Fig.\ref{fig:fig1}] and maximal mixing. For the
         purposes of example, the Higgs mass parameter $M_A$ is set to
         $2 \mu / \sin 2 \beta$. Heavy scalar, pseudoscalar and charged
         states are nearly mass degenerate: $M_{H_3}\simeq M_{A_2}
         \simeq M_{H^\pm}\simeq M_A$.}
\label{fig:higgsmass}
\end{figure}

Decays involving Higgs bosons can be quite different for different
Higgs boson mass spectra. Following the procedure outlined in Ref.\cite{mnz}
we decompose the neutral Higgs states into real and imaginary parts as
follows:
\begin{eqnarray}
H_d^0 &=& \frac{1}{\sqrt{2}}\left[v_d - S_1 s_{\beta}
          + S_2 c_{\beta} + i P_1 s_{\beta} \right], \\
H_u^0 &=& \frac{1}{\sqrt{2}}\left[v_u + S_1 c_{\beta}
          + S_2 s_{\beta} + i P_1 c_{\beta} \right], \\
S &=& \frac{1}{\sqrt{2}}\left[v_s + S_3 + i P_2 \right]
\end{eqnarray}
where the Goldstone states are removed by using the unitary gauge. We
then further rotate these states onto the mass eigenstates, $H_i$
($i=1-3$) and $A_i$ ($i=1,2$) labeled in order of ascending mass, by
using the orthogonal rotation matrices\footnote{Note that the
definitions of these mixings matrices differ slightly from those in
Ref.\cite{mnz}, where the scalar rotation matrix is defined via
$O^T = O^H$ and the pseudoscalar rotation is defined by a
rotation through an angle $\theta_A$.} $O^H$ and $O^A$:
\begin{equation}
H_i = S_j O^H_{ji}, \qquad A_i = P_j O^A_{ji}
\end{equation}
The resulting mass spectrum, composed of three scalars, two
pseudocalars, and two charged Higgs bosons, is shown in
Fig.\ref{fig:higgsmass} as a function of $\mu_{\lambda}$.  For the
purposes of example, we have chosen the mass parameter $M_A$ (defined
to be the heavy pseudoscalar mass in the MSSM limit) to be $2 \mu /
\sin 2 \beta \approx 567$~GeV, setting the scale of the heavy Higgs
bosons. The lighter Higgs bosons consist of two scalars and one
pseudoscalar. The lightest scalar and pseudoscalar in our example are
predominantly singlet states, with masses set by the scale of
$\mu_{\kappa}$.\s

Generally, the width of a 2-body neutralino or chargino $\tilde \chi_i$
decay to a neutralino or chargino $\tilde \chi_j$ and a Higgs
boson $\phi_k$ ($H_k$ or $A_k$) is given by
\begin{eqnarray}
\Gamma[{\tilde \chi}_i \to
{\tilde \chi}_j \phi_k]
&=& \frac{\lambda_{\rm PS}^{1/2}}{16 \pi m_{{\tilde \chi}_i}} \left\{
\left( m_{{\tilde \chi}_i}^2 + m_{{\tilde \chi}_j}^2 - m_{\phi_k}^2 \right)
\left( |C^L_{ijk}|^2+|C^R_{ijk}|^2 \right) \right. \nonumber \\
&& \hspace{3cm} \left. +2 \eta_{\phi} m_{{\tilde \chi}_i} m_{{\tilde \chi}_j}
\left( C^L_{ijk} C^{R \, *}_{ijk} + C^{L \, *}_{ijk} C^{R}_{ijk} \right) \right\}
\label{eq:genwidth}
\end{eqnarray}
where $\lambda_{\rm PS} = \lambda_{\rm PS}(1,m_{{\tilde \chi}_j}^2/m_{{\tilde
\chi}_i}^2, m_{\phi_k}^2/m_{{\tilde \chi}_i}^2)$ and the left/right
couplings $C^{L/R}_{ijk}$ must be specified in each individual case;
$\eta_{\phi}=1$ for $\phi=H_k,\, H^{\pm}$, and $-1$ for
$\phi=A_k$.\\ \vspace*{-2.5mm}

{\bf (i)} For the decay of a neutralino $\tilde \chi^0_i$ to a neutralino
$\tilde \chi^0_j$ and a scalar Higgs boson $H_k$, \mbox{${\tilde \chi^0}_i \to
{\tilde \chi^0}_j H_k$,} the couplings are
given by,
\begin{eqnarray}
C^R_{ijk}( {\tilde \chi^0}_i \to {\tilde \chi^0}_j H_k )
 \!&=&\! \frac{1}{2} \left[g(N^5_{i2}\!-\!N^5_{i1} t_W)
   (\phantom{-} N^5_{j3} s_{\beta}\! +\! N^5_{j4} c_{\beta})
  \!+\!\sqrt{2} \lambda
   \left( N^5_{i3} c_{\beta}\! -\! N^5_{i4} s_{\beta} \right)
         N^5_{j5}\right]\! O^H_{1k} \label{eq:sij} \nonumber \\
 \!&+&\! \frac{1}{2} \left[g(N^5_{i2}\!-\!N^5_{i1} t_W)
   (-N^5_{j3} c_{\beta}\! +\!N^5_{j4} s_{\beta})
  \!+\!\sqrt{2} \lambda \left( N^5_{i3} s_{\beta}
    \! +\! N^5_{i4} c_{\beta} \right)
    N^5_{j5} \right]\! O^H_{2k} \nonumber \\
 &+& \frac{1}{\sqrt{2}} \left[ \lambda N^5_{i3} N^5_{j4}
     -\kappa N^5_{i5} N^5_{j5} \right]\! O^H_{3k}
     \quad + \quad (i \leftrightarrow j)\\[1mm]
C^L_{ijk}( {\tilde \chi^0}_i \to {\tilde \chi^0}_j H_k )
&=&C^{R \, *}_{ijk}( {\tilde \chi^0}_i \to {\tilde \chi^0}_j H_k )
\end{eqnarray}
While the first term in each of the two square brackets in
Eq.(\ref{eq:sij}) are reminiscent of the MSSM couplings $\tilde{\chi}_i^0
{\tilde \chi}_j^0 h$ and $\tilde{\chi}_i^0 {\tilde \chi}_j^0 H$ respectively,
the other terms are genuinely new in origin, arising from the extra interaction
terms in the NMSSM superpotential. \s

\begin{figure}[t]
\vskip 0.3cm
\begin{center}
\includegraphics[height=6.8cm,clip=true]{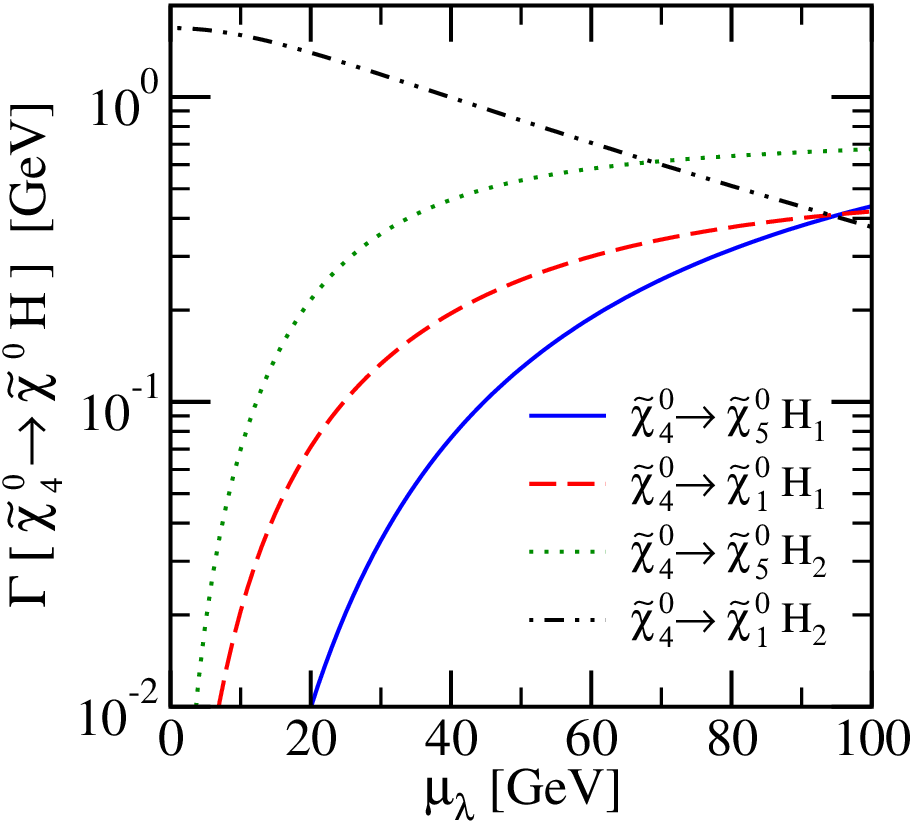} \hspace{0.3cm}
\includegraphics[height=6.8cm,clip=true]{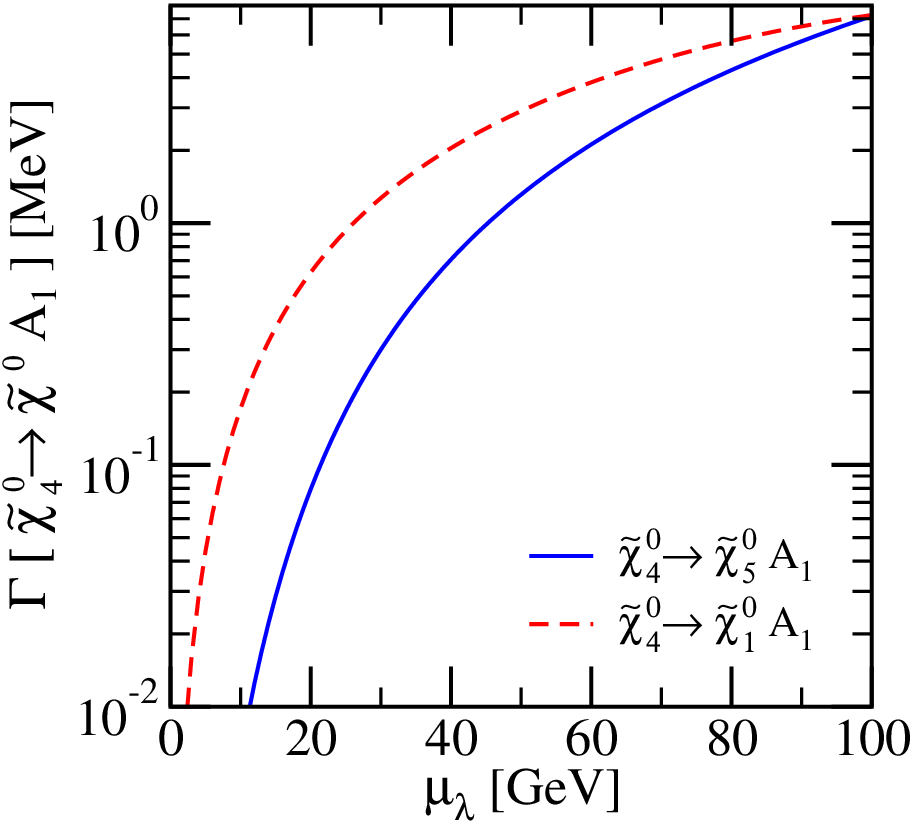}
\end{center}
\vskip -0.3cm
\caption{\it The decay widths for $\tilde \chi_4^0 \to \tilde
             \chi_{1,\,5}^0 H_{1,\,2}$ (left) and $\tilde \chi_4^0 \to
             \tilde \chi_{1,\,5}^0 A_1$ (right) for the parameter set
             $\mathbb{P}$[Fig.\ref{fig:fig1}] and maximal mixing. For
             the purposes of example, the Higgs mass parameter $M_A$
             is set to $2 \mu / \sin 2 \beta$. }
\label{fig:chitochiH}
\end{figure}

The widths for the kinematically allowed decays $\tilde \chi_4^0 \to
\tilde \chi_{1,5}^0 H_{1,2}$ are shown in Fig.\ref{fig:chitochiH} (left) as a
function of $\mu_{\lambda}$. For $\mu_{\lambda}=0$ the $\tilde \chi_5^0$ state
is decoupled from
the other neutralinos; as $\mu_{\lambda}$ is switched on, the
coupling, and therefore the decay widths, increase. The decay widths
for $\tilde \chi_4^0 \to \tilde \chi_5^0 H_{1}$ and $\tilde \chi_4^0
\to \tilde \chi_5^0 H_{2}$ are comparable, within an order of magnitude,
due to the large $\tilde{\chi}_4^0$ mass and the near mass degeneracy of
$H_1$ and $H_2$. With partial widths of order GeV, these decay modes are in
the observable range of branching ratios.\\
\vspace*{-2.5mm}

{\bf (ii)} Similarly, a 2-body neutralino decay to a neutralino and a
pseudoscalar Higgs boson, ${\tilde \chi}_i^0 \to {\tilde \chi}_j^0
A_k$, follows Eq.(\ref{eq:genwidth}) with the left/right couplings given
by
\begin{eqnarray}
C^R_{ijk} ( {\tilde \chi^0}_i \to {\tilde \chi^0}_j A_k )
 \!&=&\! \frac{1}{2} \left[ g(N^5_{i2}\!-\!N^5_{i1} t_W)
     (-N^5_{j3} s_{\beta}\! +\! N^5_{j4} c_{\beta})
    \!+\!\sqrt{2} \lambda \left( N^5_{i3} c_{\beta}\!
       +\! N^5_{i4} s_{\beta} \right) N^5_{j5} \right]\! O^A_{1k} \nonumber \\
 &+& \frac{1}{\sqrt{2}} \left[ \lambda N^5_{i3} N^5_{j4} -\kappa N^5_{i5} N^5_{j5} \right] O^A_{2k}
     \quad + \quad (i \leftrightarrow j) \label{eq:pij} \\
C^L_{ijk}( {\tilde \chi^0}_i \to {\tilde \chi^0}_j A_k )
 &=& C^{R \, *}_{ijk}( {\tilde \chi^0}_i \to {\tilde \chi^0}_j A_k )
\end{eqnarray}
Again, only the first term in the square brackets is similar to the
MSSM coupling $\bar {\tilde \chi}_i^0 {\tilde \chi}_j^0 A$. \s

The widths for the kinematically allowed decays $\tilde{\chi}^0_4\rightarrow
\tilde{\chi}^0_{1,5} A_1$ are shown in
Fig.\ref{fig:chitochiH} (right) as a function of $\mu_{\lambda}$ for
our chosen example scenario.  In comparison with the scalar case, many
of the decays are kinematically disallowed, only leaving the decays of
the heaviest two neutralinos to ${\tilde \chi}_5^0$ and the lightest
pseudoscalar ($A_1$). Note that pseudoscalar decays are strongly
suppressed compared with the scalar modes and may not be observed easily.
\\ \vspace*{-2.5mm}

{\bf (iii)} For completeness, we describe the decays of charginos to a
neutralino and charged Higgs boson ${\tilde \chi}_i^{\pm} \to {\tilde
\chi}_j^0 H^{\pm}$ ($i=1,2; j=1-5$). These follow a similar pattern,
now with the last index of the coupling removed:
\begin{eqnarray}
C^L_{ij}( {\tilde \chi}_i^{\pm} \to {\tilde \chi}_j^0 H^{\pm})
 \!\!\!&=&\!\!\! -g c_\beta \left[ N^{5 \, *}_{i4} U^*_{L \, j1}
     \!+\! \frac{1}{\sqrt{2}} \left(N^{5 \, *}_{i2}
     \!+\! N^{5 \, *}_{i1} t_W \right)U^*_{L \, j2} \right]
     \!-\! \lambda s_\beta N^{5 \, *}_{i5} U^*_{L \, j2} \\
C^R_{ij}( {\tilde \chi}_i^{\pm} \to {\tilde \chi}_j^0 H^{\pm})
 \!\!\!&=&\!\!\! -g s_\beta \left[ N^{5 \, \phantom{*}}_{i3} U^*_{R \, j1}
     \!-\! \frac{1}{\sqrt{2}} \left(N^{5 \, \phantom{*}}_{i2}
     \!+\! N^{5 \, \phantom{*}}_{i1} t_W\right)
       U^*_{R \, j2} \right]
     \!-\! \lambda c_\beta N^{5 \, \phantom{*}}_{i5} U^*_{R \, j2}
\label{eq:cij}
\end{eqnarray}
However, the large mass of the charged Higgs boson means that these
2-body decays are kinematically disallowed for our specific parameter
choice. \\ \vspace*{-2.5mm}

\begin{figure}[tbh!]
\vskip 0.3cm
\begin{center}
\includegraphics[height=6.5cm,clip=true]{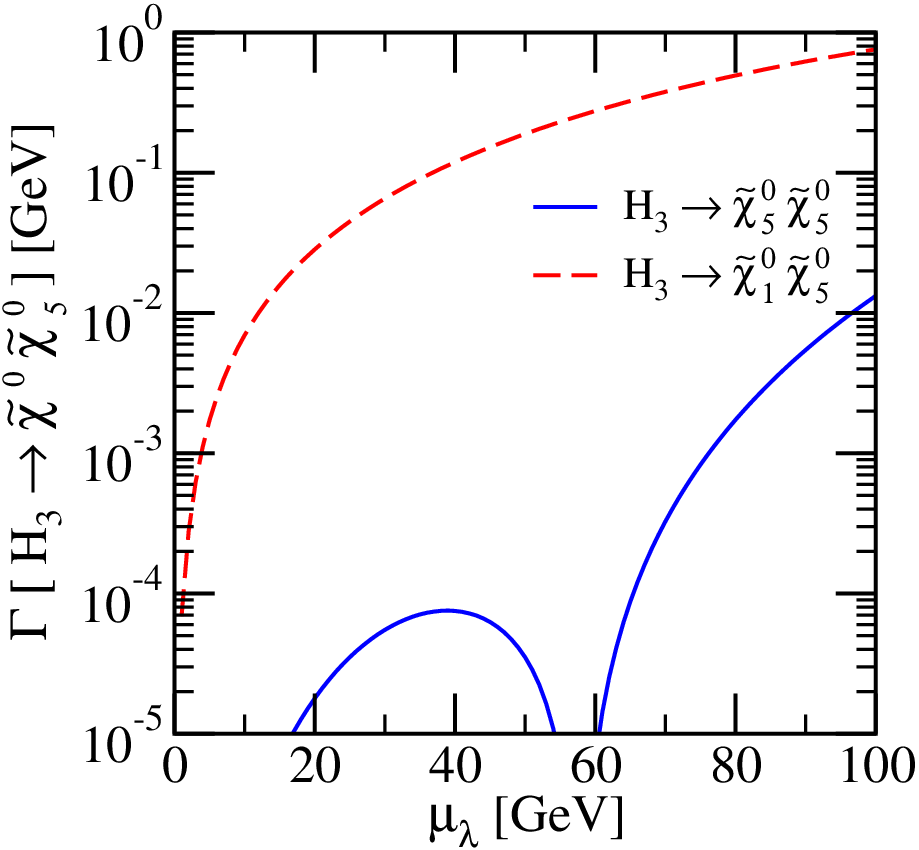} \hspace{0.5cm}
\includegraphics[height=6.5cm,clip=true]{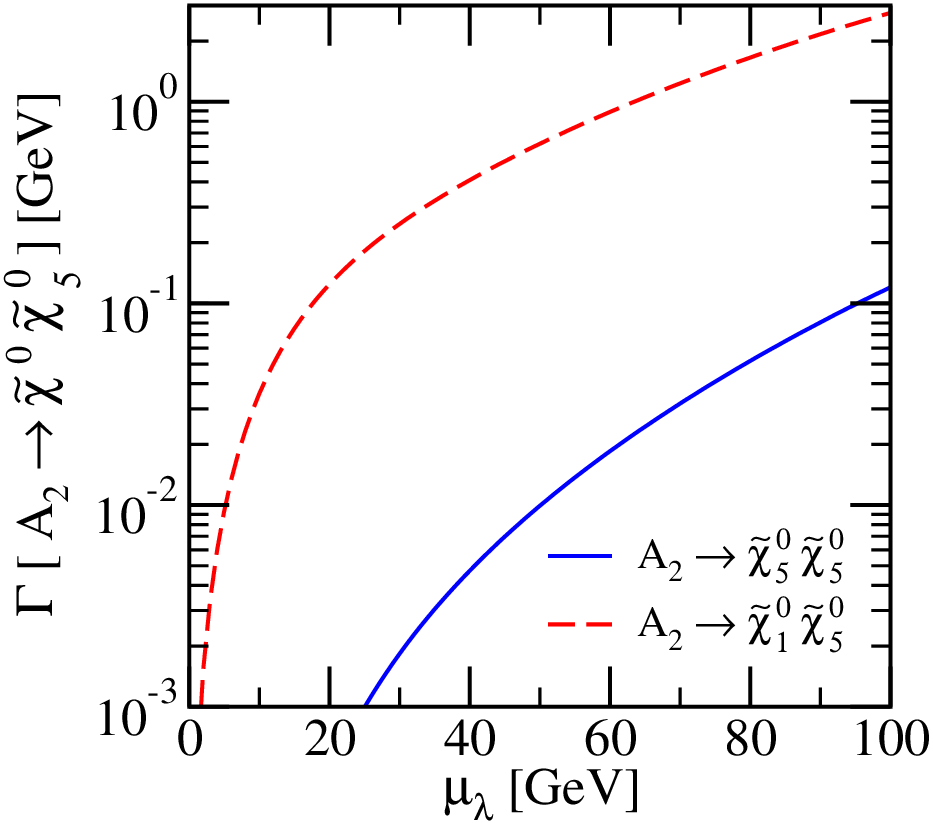} \\
\includegraphics[height=6.5cm,clip=true]{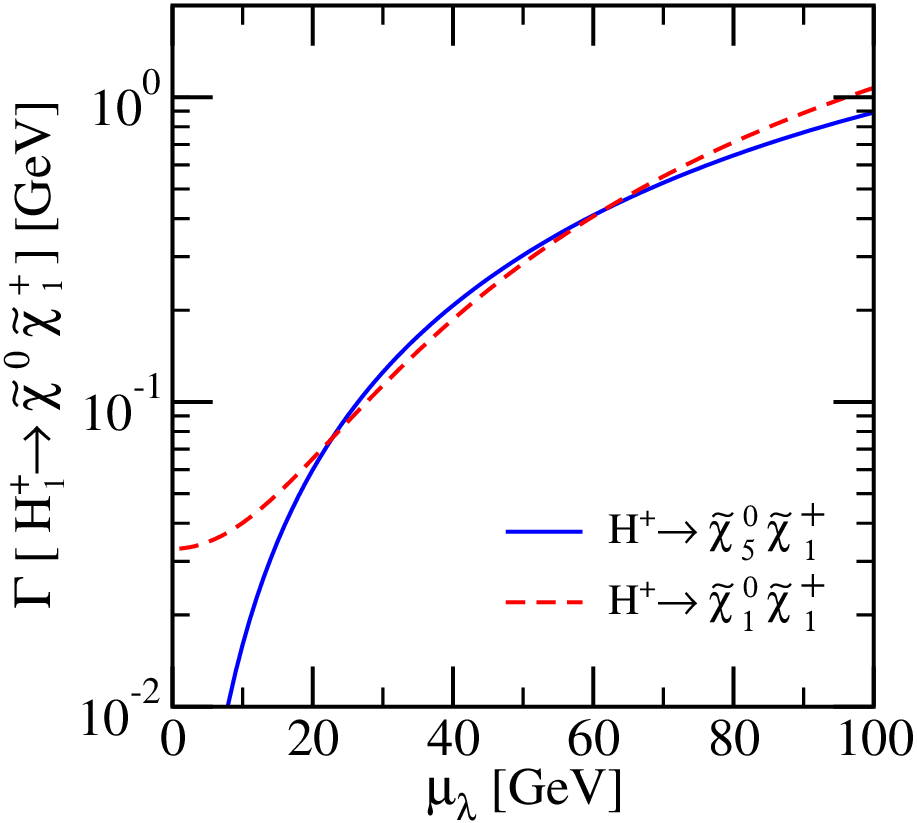}
\end{center}
\vskip -0.3cm
\caption{\it The decay widths for $H_3 \to \tilde{\chi}_{1,\,5}^0
             \tilde{\chi}_5^0$ (upper left), $A_2 \to
             \tilde{\chi}_{1,\,5}^0 \tilde{\chi}_5^0$ (upper right) and
             $H^{\pm} \to \tilde{\chi}_1^{\pm} \tilde{\chi}_{1,\,5}^{0}$
             (lower) for the parameter set
             $\mathbb{P}$[Fig.\ref{fig:fig1}] and maximal mixing. For
             the purposes of example, the Higgs mass parameter $M_A$
             is set to $2 \mu / \sin 2 \beta$. }
\label{fig:Htochichi}
\end{figure}

{\bf (iv)} It is also possible for Higgs bosons themselves to decay into the
singlino--dominated state, via the decays $H_i \to {\tilde \chi}^0_5
{\tilde \chi}^0_j$, $A_i \to {\tilde \chi}^0_5 {\tilde \chi}^0_j$ and
$H^{\pm} \to {\tilde \chi}^0_5 {\tilde \chi}^{\pm}_i$, if
kinematically allowed. Clearly this is only possible for the heavier
Higgs states; the lightest Higgs boson is never heavy enough to decay
in this way. The general form of the width for these decays $\phi_i
\to {\tilde \chi}_j {\tilde \chi}_k$ ($\phi_i = H_i,\,A_i,\,H^{\pm}$), is
given by the crossing of Eq.(\ref{eq:genwidth}):
\begin{eqnarray}
\Gamma[\phi_i \to {\tilde \chi}_j {\tilde \chi}_k]
  &=& {\cal{S}}_{jk} \frac{\lambda_{\rm PS}^{1/2}}{16 \pi m_{\phi_i}}
      \left\{\left( m_{\phi_i}^2 - m_{{\tilde \chi}_j}^2
      - m_{{\tilde \chi}_k}^2 \right)
      \left( |C^L_{ijk}|^2+|C^R_{ijk}|^2 \right) \right. \nonumber \\
  && \hspace*{3cm} \left. -2 \eta_{\phi} m_{{\tilde \chi}_j} m_{{\tilde \chi}_k}
     \left( C^L_{ijk} C^{R \, *}_{ijk} + C^{L \, *}_{ijk} C^{R}_{ijk} \right)
     \right\}
\label{eq:genwidth2}
\end{eqnarray}
where $\lambda_{\rm PS} = \lambda_{\rm PS}(1,m_{{\tilde
\chi}_j}^2/m_{\phi_i}^2, m_{{\tilde \chi}_j}^2/m_{\phi_i}^2)$ and
${\cal{S}}_{jk}=1$ or $1/2$ is the usual statistical factor.  Again,
$\eta_{\phi}=1$ for $\phi=H_k,\, H^{\pm}$, and $-1$ for $\phi=A_k$.
The couplings $C^{L/R}_{ijk}$ are related to their neutralino decay
counterparts in the obvious way:
\begin{eqnarray}
    C^{L/R}_{ijk} ( H_i \to {\tilde \chi^0}_j {\tilde \chi^0}_k)
&=& C^{L/R}_{kij} ( {\tilde \chi^0}_j \to {\tilde \chi^0}_k H_i ) \\
    C^{L/R}_{ijk} ( A_i \to {\tilde \chi^0}_j {\tilde \chi^0}_k)
&=& C^{L/R}_{kij} ( {\tilde \chi^0}_j \to {\tilde \chi^0}_k A_i ) \\
    C^{L/R}_{ij} ( H^{\pm} \to {\tilde \chi^0}_i {\tilde \chi^{\pm}}_j)
&=& C^{L/R}_{ij} ( {\tilde \chi^{\pm}}_i \to {\tilde \chi^0}_j H^{\pm} )
\end{eqnarray}
Some of these decays widths are plotted in Fig.\ref{fig:Htochichi}.
Note that a significant fraction of the Higgs boson $H_3$ and $A_2$
decays go into the invisible channel $\tilde \chi_5^0 \tilde \chi_5^0$
only if the partial decay width exceeds the range of $\sim 1/10$
GeV. The upper left panel, showing the partial width for the decay
$H_3 \to \tilde \chi_i^0 \tilde \chi_5^0$ $[i=1,5]$, has been allowed
to extend down to widths of order $10^{-5}$~GeV to show the switching
off of the $\tilde \chi_1^0 \tilde \chi_5^0 H_3$ coupling at $\sim
58$~GeV. This is caused by destructive interference between the
different constituent fields in both the Higgs and the neutralinos,
and is directly analogous to the cancellations seen in
Ref.\cite{mnz}. \s

%

%======================================
\section{Summary and Conclusions}
\label{sec:conclusion}
%======================================

In this study, we have investigated the neutralino sector of the
NMSSM, suggested by many GUT and superstring models. Moreover, this
model attempts to explain the $\mu$-problem of the MSSM by introducing
a new iso--singlet Higgs superfield, $\hat{S}$, the scalar component of
which acquires a non-zero vacuum expectation value. \s

We have given expressions for the new $5\times 5$ neutralino mass matrices
and mixing matrices and we have presented, besides the numerical analyses,
approximate analytical solutions for the neutralino masses and mixings
which provide a nice insight into the structure of the spectrum and the
mass hierarchies in case of small couplings between the MSSM and the new
iso--singlet.\s

The renormalization group flow of the parameters $\lambda$ and
$\kappa$ from the GUT scale down to the electroweak scale gives rise to
strong upper bounds on their values at the electroweak scale, where
small $\kappa$ is favored.  The qualitative features of the neutralino
masses are dependent on how strongly the PQ symmetry of the model is
broken by non-zero $\kappa$ values; this is quite accurately described by the
approximate analytical solutions.\s

If the PQ symmetry is slightly broken for small $\kappa$, the
qualitative pattern for the particle spectrum remains intact, except
that the lightest singlino-dominant neutralino acquires a mass of the
order of the electroweak scale.  Thus the model contains four MSSM--type
heavy gaugino/higgsino dominant states and one light singlino dominant
state.  Since the couplings to the $Z$ boson can be very much reduced,
the NMSSM with a slightly broken PQ symmetry constitutes a valid
scenario.\s

In contrast, a strongly broken PQ symmetry, though disfavored by the
flow of the couplings from the GUT scale down to the electroweak
scale, could provide an extra moderately heavy neutralino state,
which is only weakly coupled to the $Z$ and (s)fermions. Such decoupled
scenarios would be more difficult to distinguish the NMSSM from
the MSSM.\s

\subsection*{Acknowledgments}

The work of S.Y.C. was supported in part by
the Korea Research Foundation Grant (KRF-2002-070-C00022) and in part
by KOSEF through CHEP at Kyungpook National University.

\vspace{0.2cm}

%=============
%\appendix
%=============

\section*{Appendix: The small--mixing approximation}

The $5\times 5$ neutralino mass matrix of Eq.(\ref{eq:mass_matrix})
in general cannot be diagonalized analytically to derive the physical
neutralino masses. However, simple analytical expressions for masses
and mixing parameters can be found by making use of approximations for
small doublet--singlet mixing which is theoretically very well motivated.\s

To construct this approximate solution in the neutralino sector, we
treat the doublet--singlet mixing parameter $\mu_\lambda \ll M_i, \mu$,
together with the $Z$--boson mass $m_Z$, as small parameters of generic
size $\varepsilon \ll 1$ in units of the typical SUSY masses. Then,
as long as these SUSY masses
%and their differences from $\mu_\kappa$
are not as small as $\mu_\lambda$, we observe a hierarchical structure
in the neutralino mass matrix of the form:
\begin{eqnarray}
{\mathcal H} =
  \left(\begin{array}{cc}
    A         &   X \\
   X^T  &    B
        \end{array}\right)
\end{eqnarray}
where $A$ is a $4\times 4$ matrix incorporating elements of the order of
the large SUSY scale, $B$ is a scalar and $X$ is a 4-component vector of
order $\varepsilon$.

Performing an auxiliary orthogonal transformation $O$ defined by
the matrix\footnote{Note that by standard notation $\Omega  \Omega^T$ is
a $4\times 4$ matrix with the elements $\Omega_i \Omega_j$ while
$\Omega^T \Omega$ is a scalar with the value
$\sum\, \Omega^2_i$.},
\begin{eqnarray}
 O =
  \left(\begin{array}{cc}
    1 \hspace{-0.14cm} 1_{4\times 4} - \frac{1}{2}\, \Omega \Omega^T
  &  \Omega \\[2mm]
    -\Omega^T & 1
   -\frac{1}{2}\, \Omega^T \Omega
\end{array} \right)
\label{eq:othogonal_matrix}
\end{eqnarray}
with the mixing column vector $\Omega = [A-B]^{-1} X$, the
mass matrix takes block diagonal form, accurate to order
$\varepsilon^2$:
\begin{eqnarray}
O {\mathcal H}\, O^{T} &=& \left(
\begin{array}{cc}
A + \Delta_A
   &  0\\[1mm]
0  &  B+\Delta_B
\end{array} \right)
\label{eq:block_diagonal_general}
\end{eqnarray}
where
\begin{eqnarray}
&& \Delta_A = \frac{1}{2}\left\{{[A-B]}^{-1},\, XX^T \right\}
               \nonumber\\[1mm]
&& \Delta_B = -X^T {[A-B]}^{-1}X
\end{eqnarray}
Both $\Delta$'s are of order $\varepsilon^2$.
If $A$ is diagonal, only the diagonal elements of $\Delta_A$
need be kept as re--diagonalization would change the mass matrix
(\ref{eq:block_diagonal_general}) and the orthogonal matrix
(\ref{eq:othogonal_matrix}) only beyond the order considered in the
systematic expansion. We note that the correction terms satisfy the simple
sum rule ${\rm Tr}{\Delta_A}+\Delta_B=0$.\s

If $B$ is also as small as the elements of the low vector\, $X$,
the mixing column vector $\Omega = A^{-1} X$ and the
correction terms $\Delta_A$ and $\Delta_B$ are further simplified to be
\begin{eqnarray}
\Delta_A = \frac{1}{2}\left\{A^{-1},\, XX^T \right\}\ \ {\rm and}\ \
\Delta_B = -X^T A^{-1} X
\end{eqnarray}
On the contrary, if $B$ is much larger than the other parameters,  the mixing
column vector $\Omega = -X/B$ and the correction terms take the following
simple form
\begin{eqnarray}
\Delta_A = -XX^T/B,\qquad
\Delta_B =  X^TX/B
\end{eqnarray}
Both these approximations have been used in the derivation of all the mass and
mixing formulae discussed earlier in the report.\s

The diagonalization of the mass matrix ${\cal M}_5$,
\begin{eqnarray}
{\cal M}_5 = \left(\begin{array}{cc}
          {\cal M}  &  X  \\
             X^T    &  \tilde{m}_5
             \end{array}\right)
\end{eqnarray}
with ${\cal M}$ being the $4\times 4$ MSSM  mass sub--matrix,
makes use of the
block--diagonalization method in the following way:\s

\noindent
{\bf (1)} In the first step ${\cal M}$ is diagonalized by applying the
          well--elaborated MSSM procedure
\begin{eqnarray}
{\cal M}^D = V{\cal M} V^T
\end{eqnarray}
generating the eigenvalues ${\cal M}^D={\rm diag}\left[\tilde{m}_1,\ldots,
\tilde{m}_4\right]$. \s

\noindent
{\bf (2)} The ensuing $5\times 5$ matrix can subsequently be block--diagonalized
          as worked out above by applying the orthogonal transformation in
          Eq.(\ref{eq:othogonal_matrix}) with
\begin{eqnarray}
\Omega = V\,\Gamma \;\;:\;\;
\Gamma = V^T \left({\cal M}^D - \tilde{m}_5\right)^{-1} V X
       = \left({\cal M} - \tilde{m}_5\right)^{-1} X
\end{eqnarray}
Note that $\Gamma$ is of order $\varepsilon$ -- quantum mechanically
enhanced however if mass differences $|\tilde{m}_i - \tilde{m}_5|$ are,
accidentally, small.  \s

\noindent
{\bf (3)} The block--diagonalization affects the upper left diagonalized
          $4\times 4$ submatrix ${\cal M}^D$ only beyond second order and
          likewise the orthogonal matrix $V^5$ beyond the second and first
          order considered, respectively, for on-- and off--diagonal elements.
          As a result, we obtain the final diagonal form of the mass matrix
          as
\begin{eqnarray}
{\cal M}^D_5 \approx V^5 {\cal M}_5 V^{5T}
\end{eqnarray}
with
\begin{eqnarray}
V^5\, \approx\,
 \left(\begin{array}{cc}
    1 \hspace{-0.14cm} 1_{4\times 4}
    - \frac{1}{2}(V\Gamma) (V\Gamma)^T
  &  (V\Gamma) \\[2mm]
    -(V\Gamma)^T & 1
   -\frac{1}{2} (V\Gamma)^T (V\Gamma)
      \end{array} \right)\,
    \left(\begin{array}{cc}
          V  &  0 \\[2mm]
          0  &  1 \hspace{-0.14cm} 1_{1\times 1}
        \end{array}\right)
\end{eqnarray}
in obvious notation. While the right-most part solves the MSSM
diagonalization, the left-most part diagonalizes the NMSSM under the
assumption of small doublet--singlet mixing.\s

\end{document}